# Early evolution of the solar accretion disk inferred from Cr-Ti-O isotopes in individual chondrules

Jonas M. Schneider[1]*, Christoph Burkhardt[1], Yves Marrocchi[2], Gregory A. Brennecka[3], and Thorsten Kleine[1]

*corresponding author: jonas.m.schneider@uni-muenster.de

[1]Institut für Planetologie, University of Münster, Wilhelm-Klemm-Straße 10, 48149 Germany,

[2]CRPG, CNRS, Université de Lorraine, UMR 7358, Vandoeuvre-lès-Nancy, 54501, France

[3]Nuclear and Chemical Sciences Division, Lawrence Livermore National Laboratory, Livermore, CA 94550, US

abstract 292 words

main text 6392 words

1 Table

5 Figures

Supplementary Materials: 1 Table, 1Figure

**Abstract**

Isotopic anomalies in chondrules hold important clues about the dynamics of mixing and transport processes in the solar accretion disk. The meaning of these anomalies is debated and they have been interpreted to indicate either disk-wide transport of chondrules or local heterogeneities of chondrule precursors. However, all previous studies relied on isotopic data for a single element (either Cr, Ti, or O), which does not allow distinguishing between source and precursor signatures as the cause of the chondrules' isotope anomalies. To overcome this problem, we obtained the first combined O, Ti, and Cr isotope data for individual chondrules from enstatite, ordinary, and carbonaceous chondrites. We find that chondrules from non-carbonaceous (NC) chondrites have relatively homogeneous $\Delta^{17}O$, $\varepsilon^{50}Ti$, and $\varepsilon^{54}Cr$, which are similar to the compositions of their host chondrites. By contrast, chondrules from carbonaceous chondrites (CC) have more variable compositions, some of which differ from the host chondrite compositions. Although the compositions of the analyzed CC and NC chondrules may overlap for either $\varepsilon^{50}Ti$, $\varepsilon^{54}Cr$, or $\Delta^{17}O$, in multi-isotope space none of the CC chondrules plot in the compositional field of NC chondrites, and no NC chondrule plots within the field of CC chondrites. As such, our data reveal a fundamental isotopic difference between NC and CC chondrules, which is inconsistent with a disk-wide transport of chondrules across and between the NC and CC reservoirs. Instead, the isotopic variations among CC chondrules reflect local precursor heterogeneities, which most likely result from mixing between NC-like dust and a chemically diverse dust component that was isotopically similar to CAIs and AOAs. The same mixing processes, but on a larger, disk-wide scale, were likely responsible for establishing the distinct isotopic compositions of the NC and CC reservoirs, which represent in inner and outer disk, respectively.

# 1. Introduction

Chondritic meteorites provide fossilized remnants of the materials present in the early Solar System, and are key samples to unravel the early dynamical evolution of the solar accretion disk (Scott and Krot, 2013). Chondrites are unequilibrated assemblages of various low- and high-temperature components, including volatile-rich, fine-grained matrix, Fe-Ni metal, refractory inclusions, and the eponymous chondrules, which constitute up to 90% of bulk chondrites (e.g., Scott, 2007). Although the conditions, process(es), and chronology of chondrule formation remain debated (for a review see Connolly and Jones, 2016), their ubiquitous presence in the disk makes them a prime tracer of transport, mixing, and accretion processes in the early Solar System.

Utilizing chondrules in this manner requires understanding the genetic relationships among individual chondrules, and between chondrules and their host chondrite. This information may be obtained from isotope anomalies in individual chondrules. Of these, O isotopes are the most extensively studied, but the O isotope variability among chondrules likely represents a combination of precursor heterogeneity, open-system exchange with the nebular gas, and fluid-assisted alteration on the parent body (e.g., Clayton, 2003; Kita et al., 2010; Piralla et al., 2020). As such, the genetic heritage of chondrules can be difficult to interpret using O isotopes alone.

Other isotope tracers that have been used to assess the genetic heritage of chondrules include variations in the relative abundance of $^{50}$Ti and $^{54}$Cr. Anomalies in these two isotopes are nucleosynthetic in origin and are powerful tools to investigate genetic relationships between individual solids, bulk meteorites, and planets (e.g., Gerber et al., 2017; Trinquier et al., 2009; Warren, 2011). For instance, isotopic anomalies in $^{50}$Ti and $^{54}$Cr, together with O isotope variations, were instrumental in identifying a fundamental isotopic dichotomy between non-carbonaceous (NC) and carbonaceous (CC) meteorites (Warren, 2011). Subsequent work has shown that this dichotomy extends to several other elements, most notably Mo (Budde et al., 2016a; Poole et al., 2017; Worsham et al., 2017). The NC and CC meteorites represent two genetically distinct disk reservoirs that remained spatially separated for several millions of years (Ma), most likely through the early formation of Jupiter (Kruijer et al., 2017). The NC reservoir is therefore thought to represent the primordial inner, and the CC reservoir the primordial outer Solar System.

The existence of the NC-CC dichotomy among bulk meteorites raises the question of whether this dichotomy also exists at the individual chondrule scale. This might be expected, given that the dichotomy was likely caused by distinct dust compositions between the two reservoirs. As chondrules provide a snapshot of the dust composition at a particular time and location in the disk, chondrules from the NC and CC reservoir should have fundamentally different isotopic compositions, at least for elements that define the NC-CC dichotomy. However, the isotopic composition of chondrules may also vary because the chondrules themselves derive from different areas of the disk and were subsequently transported to the accretion region of their host chondrite. Currently available Cr and Ti isotopic data do not allow distinguishing between these two disparate interpretations and have been interpreted both as source signatures (Olsen et al., 2016) and local heterogeneities among chondrule precursors (Gerber et al., 2017).

A major problem common to previous isotopic studies on individual chondrules is that the isotopic data were only obtained for a single element. Although such studies reveal isotopic heterogeneities among chondrules for each investigated element, the interpretation of a single isotopic signature in terms of large-scale disk processes remains inherently uncertain. To overcome this shortcoming, we obtained combined Cr, Ti, and O isotopic data for a comprehensive set of individual NC and CC chondrules. These data provide new and firm constraints on the origin of isotope anomalies in chondrules and their relation to the NC-CC dichotomy observed among bulk meteorites. Combined, these new constrains have important implications for understanding the early evolution of the solar accretion disk.

## 2. Material and Methods

### 2.1 Samples

All samples of this study were part of the Ti isotope study by Gerber et al. (2017) and include single chondrules as well as bulk rocks from the carbonaceous chondrites Allende (CV3), GRA06100 (CR2), and NWA081 (CR2), the enstatite chondrite MAC02837 (EL3) and the ordinary chondrite Ragland (LL3.4). For this study, seven chondrules from Allende, six from GRA06100, one from NWA081, nine from Ragland, and four from MAC02837 were selected. In addition, matrix and pooled Allende chondrule separates, consisting of hundreds of chondrules and chondrule fragments (Budde et al., 2016a), were also analyzed.

After gently crushing pieces of bulk chondrite in an agate mortar, individual chondrules were hand-picked with tweezers under a binocular microscope. Whenever possible, whole, spherical chondrules were selected. Adhering dust was removed from chondrule rims by sonication in acetone for 15–30 min. The individual chondrules were photographed with a Keyence VHX-500F digital microscope (SM Fig. S1), and then broken in an agate mortar. A fragment comprising between 5-20 wt.% of the bulk chondrule was embedded into epoxy for petrographic characterization by SEM (Fig. S1) and in-situ O isotopic analyses by SIMS. The remaining chondrule material, comprising 80-95 wt.% of the original bulk chondrule, was dissolved in purified mineral acids, and Ti was separated by ion exchange chromatography (Gerber et al. 2017). The Cr isotope measurements of this study were made on purified Cr from the column washes of the Ti chemistry of Gerber et al. (2017). Chromium concentrations were determined on the column washes prior to Cr separation by ion exchange chromatography, using a ThermoScientific X-Series II quadrupole inductively coupled plasma-mass spectrometer (ICP-MS).

*2.2 Oxygen isotope measurements by SIMS*

Thin sections of the chondrule fragments were examined microscopically in transmitted and reflected light. Scanning electron microscope observations were performed at CRPG-CNRS (Nancy, France) using a JEOL JSM-6510 with 3 nA primary beam at 15 kV. Cathodoluminescence (CL) imaging of chondrules was performed using an RGB CL-detection unit attached to a field emission gun secondary electron microscope JEOL J7600F at the Service Commun de Microscopie Électronique (SCMEM, Nancy, France). We measured the O isotopic compositions of chemically characterized olivine crystals with a CAMECA ims 1270 E7 at CRPG-CNRS. $^{16}O^-$, $^{17}O^-$, and $^{18}O^-$ ions produced by a $Cs^+$ primary ion beam (~15 µm, ~4 nA) were measured in multi-collection mode with two off-axis Faraday cups (FCs) for $^{16,18}O^-$ and the axial FC for $^{17}O^-$ (see Marrocchi et al., 2018a). To remove $^{16}OH^-$ interference on the $^{17}O^-$ peak and to maximize the flatness atop the $^{16}O^-$ and $^{18}O^-$ peaks, the entrance and exit slits of the central FC were adjusted to obtain a mass resolution power (MRP) of ~7000 for $^{17}O^-$. The multi-collection FCs were set on slit 1 (MRP = 2500). Total measurement times were 240 s (180 s measurement + 60 s pre-sputtering). We used three terrestrial standard materials (San Carlos olivine, magnetite, and diopside) to define the instrumental mass fractionation line for the three oxygen isotopes and correct for instrumental mass fractionation due to the matrix effect of olivine. To increase the precision on the measurements, we consecutively analyzed four standards, eight chondrule olivine crystals, and four standards. Typical count rates obtained on the San Carlos olivine standards were $2.5 \times 10^9$ cps for $^{16}O$, $1.0 \times 10^6$ cps for $^{17}O$, and $5.4 \times 10^6$ cps for $^{18}O$. Data are given in the delta notation $\delta^{17,18}O$ (=[($^{17,18}O/^{16}O_{sample}/^{17,18}O/^{16}O_{SMOW}$)-1] ×1000), and capital delta notation $\Delta^{17}O$ (=$\delta^{17}O-0.52\times\delta^{18}O$). Depending on sample homogeneity and size, between 2 and 25 individual measurements were obtained on each chondrule fragment. Only one measurement per olivine grain was carried out and each SIMS spot was then checked by SEM. Analyses from cracks or grain boundaries were not included. The 2σ errors were ~0.2‰ for $\delta^{18}O$, ~0.4‰ for $\delta^{17}O$, and 0.9‰ for $\Delta^{17}O$. The error on $\Delta^{17}O$ was calculated by quadratically summing the errors on $\delta^{17}O$ and $\delta^{18}O$ and the standard deviation of $\Delta^{17}O$ values of the four terrestrial standards.

## 2.3 Chromium separation and isotopic measurement by TIMS

Chromium was separated from the sample matrix using a three-stage ion exchange chromatography. The first two steps were similar to the procedures from Yamakawa et al. (2009) and Trinquier et al. (2008), employing both anion and cation exchange resins. The elution procedure on the third column (cation exchange resin AG50W-X8, 200-400 mesh) was slightly modified compared to the two aforementioned studies. The Cr cut from the second column was loaded in 0.5 M $HNO_3$ and residual matrix elements were rinsed off the column using 0.5 M $HNO_3$, 0.5 M HF, and 6 ml 1 M HCl, before Cr was eluted with 10 ml of 2 M HCl. To remove residual organics, the Cr cut was then treated with 2×4 drops of *aqua regia* at 120 °C, 2×2 drops of concentrated $HNO_3$ at 120 °C, and 2×1 drop of $H_2O_2$ at 80 °C. The sample was then converted with concentrated HCl twice before it was taken up in an appropriate volume of 6 M HCl to obtain a ~500 µg/g loading solution. The Cr yield of the full chemical procedure was determined on aliquots by ICP-MS, and was generally >70%. Aliquots of the NIST 3112a standard processed through the same chemistry yielded indistinguishable isotopic results as unprocessed standards.

The Cr isotope measurements were performed on a ThermoScientific Triton *Plus* thermal ionization mass spectrometer (TIMS) at the Institut für Planetologie in Münster. Approximately 1 µg Cr per sample was loaded onto single outgassed Re filaments and covered with an Al-Si-gel emitter. The emitter was made from 3 g of Aerosil 300 (Evonik) Si-Gel in 60 ml MQ, a portion of this was then mixed 1:1 with a 5 g/L $H_3BO_3$ solution and a 1000µg/g Al solution. Ion beams were simultaneously collected in static mode for all four chromium isotopes ($^{50}Cr$, $^{52}Cr$, $^{53}Cr$, and $^{54}Cr$) using Faraday cups connected to $10^{11}$ Ω feedback resistors. The signal intensity was usually ~10 V on $^{52}Cr$ and 200-250 mV on $^{54}Cr$. Possible interferences from $^{50}Ti$, $^{50}V$, and $^{54}Fe$ were monitored using $^{49}Ti$, $^{51}V$, and $^{56}Fe$ beams in cups L4, L1, and H4, respectively. However, Ti and V signals were not detected above baseline during the measurements of this study, and interferences from Fe were minimal. Measured $^{56}Fe/^{52}Cr$ ratios were typically $< 10^{-5}$, well below the $10^{-4}$ threshold for successful Fe interference correction (Qin et al., 2010). We tested the accuracy of the Fe interference correction by multiple analyses on which high amounts of Fe were loaded together with the Cr standard. Repeated measurement of individual filaments showed a decrease in the Fe signal over time, but no change in the final interference-corrected $^{54}Cr/^{52}Cr$ ratio.

Data acquisition consisted of 50 blocks of 30 cycles each (1500 total cycles), 4s integration time per cycle, and a baseline and refocus every three blocks. Instrumental mass fractionation was corrected by internal normalization to $^{50}Cr/^{52}Cr = 0.051859$ (Shields et al., 1966) using the exponential law (Russell et al., 1978). All data are reported in ε-units as the parts-per-ten-thousand deviation from the $^{53}Cr/^{52}Cr$ and $^{54}Cr/^{52}Cr$ measured for the NIST SRM3112a Cr standard in each session. Repeated measurements of the NIST SRM3112a Cr standard in each session (n = 7–14) provided an external reproducibility (2s.d.) of ±0.2 $ε^{53}Cr$ and ±0.3 $ε^{54}Cr$. Depending on the total amount of Cr available, samples were loaded on 1-3 filaments and each filament was measured multiple times if possible, resulting in up to seven individual measurements for individual chondrules. For the pooled chondrule separates, matrix fractions, and bulk Allende, the number of measurements was larger as more Cr was available for these samples. Reported uncertainties for sample measurements are 95% confidence limits of the mean (n ≥ 4), or twice the standard deviation obtained for the NIST SRM3112a Cr standard that was analyzed during the same session (n < 4).

## 3. Results

### 3.1 Petrography and O isotopes

Petrographic characteristics of the individual chondrules are provided in Table 1 and in Fig. S1. Except for one barred olivine (BO) chondrule, and one Al-rich chondrule, all investigated chondrules from the CV chondrite Allende are porphyritic olivine (PO) and olivine-pyroxene (POP) type I (low FeO) chondrules. All investigated chondrules from the CR chondrites GRA06100 and NWA081 are porphyritic olivine-pyroxene-rich (POP) type I chondrules. All but one chondrule from the ordinary chondrites Ragland are type II chondrules (75<Mg#<90) and include porphyritic chondrules (PO, POP), radial pyroxene chondrules (RP, with systematic presence of small olivine grains) and one crypto-crystalline (C) chondrule. Finally, the four investigated chondrules from the enstatite chondrite MAC02837 are type I porphyritic pyroxene (PP) and radial pyroxene (RP) chondrules. The diameters of the analyzed chondrules range from 0.7 to 1.9 mm for CC chondrules, from 1.0 to 2.7 mm for OC, and from 0.9 to 1.7 mm for EC chondrules.

The O isotopic compositions of the chondrules are provided in Table 1, and plotted in Fig. 1. The data of the individual analyses for each chondrule are given in the supplementary materials (SM, Table S1). For most chondrules, measured olivine grains of a given chondrule exhibit homogeneous $\Delta^{17}O$ within uncertainty. For these samples averaging the individual olivine measurements provides a good estimate of the bulk chondrule's O isotopic composition (Table S1). However, three CV and two CR chondrules contain relict grains, which are defined as having a $\Delta^{17}O$ value that differs from that of their host chondrule outside of 3σ (Ushikubo et al., 2012). An SEM image of the CV chondrule fragment with the most $^{16}O$-rich relict grain ($\Delta^{17}O$ = -14‰) is shown in Fig. 2, along with an CL image of the same sample. For the samples with relict grains, averaging of the olivine $\Delta^{17}O$ values provides a biased estimate of the bulk chondrule composition towards the composition of the relict grains. This is because we have not randomly sampled the chondrules, and chondrules do not consist solely of olivine. Prior studies have shown that low-Ca pyroxenes are isotopically homogeneous for O, and that they have the same O isotopic compositions as the host olivines (Tenner et al., 2018; 2015). We therefore calculated the "bulk" O-isotopic composition of chondrules having relict grains by considering the host olivines only. Importantly, the O isotopic compositions of host minerals reflects contributions from O inherited from chondrule precursors (through relict mineral dissolution) and from O coming from the gas (i.e., SiO, $H_2O$; Libourel and Portail, 2018; Marrocchi et al., 2019). Our bulk estimate, therefore, accounts for the effect of relict mineral dissolution and should be close to the actual bulk O isotopic composition (Table S1).

In an O three-isotope diagram the bulk CC chondrules plot along a slope ~1 line below the terrestrial fractionation line (TFL), whereas the NC chondrules are clustered on or above the TFL (Fig. 1). The ordinary chondrite chondrules display indistinguishable $\Delta^{17}O$ values averaging at 1.4±0.7 ‰. Similarly, the enstatite chondrite chondrules exhibit homogeneous $\Delta^{17}O$ with an average value of 0.02±0.51‰. The CR chondrules display more variable $\Delta^{17}O$ with values from -3.9 to -1.5 ‰, and for chondrules from Allende $\Delta^{17}O$ varies from -3 to -6 ‰. It is important to recognize that the uncertainty on the bulk $\Delta^{17}O$ for some of the individual CC chondrules is larger because the chondrules are internally heterogeneous. Thus, within and among individual CC chondrules (in particular CV chondrules) there is O isotope variability. By contrast, different grains within individual NC chondrules are isotopically homogeneous, and variability among NC chondrules is limited.

*3.2 Chromium isotopes*

The Cr isotope data are reported in Table 1. In addition to the chondrules and bulk chondrites, we also report data for the terrestrial rock standards BIR1a, DTS2b, and JA-2. These were processed through the full analytical procedures alongside the meteorite samples to assess the accuracy and precision of the Cr isotope measurements. The $\varepsilon^{53}$Cr values of the terrestrial samples are indistinguishable from NIST SRM3112a, but their $\varepsilon^{54}$Cr seems to be offset to slightly higher values. This is consistent with similar small offsets reported for measured $\varepsilon^{54}$Cr of terrestrial basalts in prior studies (Mougel et al., 2018; Qin et al., 2010). Similar offsets of natural rocks from standard solutions are observed for other elements (e.g., Ti, Mo) and are likely due to non-exponential fractionation processes, possibly induced during standard production (Zhang et al., 2012; Budde et al., 2019).

The Cr isotopic data for the bulk chondrite samples are also in good agreement with previously reported data for samples from the same chondrite groups (Trinquier et al. 2007; Qin et al., 2010; Göpel et al., 2015). Combined, the data for terrestrial samples and bulk chondrites therefore show good inter-laboratory agreement.

Individual ordinary chondrite chondrules display variable $\varepsilon^{53}$Cr ranging from ~0.25 to ~0.74, which for most chondrules are also higher than the bulk ordinary chondrite value. Chondrules from the other chondrite groups investigated in this study show less variability and are more similar to the values of their respective host chondrites. The variable $\varepsilon^{53}$Cr among the OC chondrules almost certainly reflect radiogenic variations resulting from the decay of short-lived $^{53}$Mn ($t_{1/2}$ ~3.7 Ma). However, as the Mn/Cr ratios of the chondrules were not determined in this study, the magnitude of potential radiogenic $\varepsilon^{53}$Cr variations cannot be quantified. Regardless, the main focus of this study is on nucleosynthetic $\varepsilon^{54}$Cr variations, and the $\varepsilon^{53}$Cr will therefore not be discussed any further.

The $\varepsilon^{54}$Cr values of the EC, OC, and CC chondrules and their host bulk chondrites are shown in Fig. 3. The OC chondrules display no resolvable $\varepsilon^{54}$Cr variations and their mean $\varepsilon^{54}$Cr is indistinguishable from the host chondrite's $\varepsilon^{54}$Cr. Likewise, the $\varepsilon^{54}$Cr of the single EC chondrule of this study is indistinguishable from the $\varepsilon^{54}$Cr of bulk enstatite chondrites. Compared to the EC and OC chondrules, all CR chondrules have elevated $\varepsilon^{54}$Cr, and most of them are indistinguishable from the $\varepsilon^{54}$Cr of bulk CR chondrites. By contrast, individual Allende chondrules exhibit highly variable $\varepsilon^{54}$Cr, with a continuum from ~0.1 to ~1.4 $\varepsilon^{54}$Cr (Fig. 3). The $\varepsilon^{54}$Cr results for CR and CV chondrules are consistent with those of Olsen et al. (2016), although these authors reported a somewhat larger range of $\varepsilon^{54}$Cr values among CR and CV chondrules (Fig. 3). Finally, the three pooled chondrule separates from Allende, each consisting of hundreds of chondrules (Budde et al., 2016b), are indistinguishable from one another and have an average $\varepsilon^{54}$Cr of 0.58±0.12. This value is resolved from the bulk CV composition ($\varepsilon^{54}$Cr ~1) and from Allende matrix, which has an average $\varepsilon^{54}$Cr of ~1.

*3.3 Titanium isotopes*

To facilitate direct comparison of the O and Cr isotopic data with previously obtained Ti isotope data on the same chondrules, the Ti concentration and $\varepsilon^{50}$Ti data from Gerber et al. (2017) are also provided in Table 1, and the results are briefly summarized here. Whereas EC and OC chondrules exhibit only small $\varepsilon^{50}$Ti variations relative to the composition of their host chondrites, CC chondrules are characterized by a broad range of $\varepsilon^{50}$Ti values, ranging from the slightly negative $\varepsilon^{50}$Ti typically observed for OC chondrules to the large $\varepsilon^{50}$Ti excesses known from Ca–Al-rich inclusions (CAIs) (Fig. 4). The CC chondrules also display a broad correlation of $\varepsilon^{50}$Ti and Ti concentration, indicating the addition of isotopically heterogeneous and Ti-rich—i.e., CAI-like—material to enstatite and ordinary chondrite-like chondrule precursors (Gerber et al., 2017).

## 4. Discussion

While the new O and Cr isotope data for single chondrules are consistent with results of previous studies, the major advantage of this study is that, for the first time, the Cr, O, and Ti isotopic data have been obtained for the same individual chondrules. We will show below that although the isotopic data for each element alone reveal isotopic heterogeneity among chondrules, only the combined data of all three elements make it possible to disentangle the process(es) responsible for these heterogeneities.

The key observations of the combined Cr, Ti, and O isotopic data may be summarized as follows (Fig. 4). First, OC and EC chondrules display relatively homogeneous $\Delta^{17}O$, $\varepsilon^{50}Ti$, and $\varepsilon^{54}Cr$, which are indistinguishable from the compositions of their respective host chondrites. Second, CC chondrules have, on average, more elevated $\varepsilon^{50}Ti$ and $\varepsilon^{54}Cr$, and lower $\Delta^{17}O$ compared to NC (i.e., OC and EC) chondrules. Although the compositions of CC and NC chondrules may overlap for either $\varepsilon^{50}Ti$, $\varepsilon^{54}Cr$, or $\Delta^{17}O$, in multi-isotope space none of the investigated CC chondrules plots in the compositional field of NC chondrites, and none of the investigated NC chondrules plots within the CC field (Fig. 4a-c). Finally, in contrast to NC chondrules, CC chondrules—and in particular CV chondrules—exhibit variable isotope anomalies for all three elements. The distinct isotopic compositions of NC and CC chondrules for a range of geo- and cosmochemically different elements, combined with the observation of isotopic homogeneity in NC chondrules versus isotopic heterogeneity in CC chondrules represents a fundamental difference between NC and CC chondrites.

Below, we first attempt to identify the carrier of the isotopic anomalies among the chondrules, and then use the combined isotopic data to evaluate mixing and transport processes before and during chondrite parent body accretion. Finally, we will place the chondrule compositions into the broader context of the evolution of distinct disk reservoirs, namely the NC and CC reservoirs.

*4.1 Origin of isotopic heterogeneities among chondrules*

When the $\varepsilon^{54}$Cr data for chondrules from NC and CC chondrites are considered together, chondrules from chondrites with a higher abundance of CAIs (i.e., CV, CO) exhibit more $\varepsilon^{54}$Cr variability compared to chondrites with lower CAI abundances (i.e., EC, OC, CR). The same observation holds for $\varepsilon^{50}$Ti variations among chondrules, which have been attributed to the heterogeneous distribution of CAIs among the chondrule precursors (Gerber et al., 2017). Specifically, the $\varepsilon^{50}$Ti variations among the CV chondrules are correlated with their Ti content, consistent with the variable addition of $^{50}$Ti- and Ti-rich CAIs to chondrule precursors (Gerber et al., 2017). By contrast, there is no obvious correlation between $\varepsilon^{54}$Cr and Cr concentration among the CV chondrules, neither in the data of this study, nor when data for CV and CO chondrules from previous studies are also included (Olsen et al., 2016; Zhu et al., 2019). The $\varepsilon^{54}$Cr anomalies in the chondrules may, therefore, reflect admixture of a isotopically highly anomalous component, in which case small amounts of admixed material are sufficient to generate the observed $\varepsilon^{54}$Cr variations, while the Cr concentrations would remain unaffected. Alternatively, the admixed material is isotopically less anomalous and had a Cr concentration similar to the chondrules themselves.

Materials known to have significant $^{54}$Cr excesses and which may be capable of shifting bulk isotopic compositions include presolar grains and CAIs. For instance, Cr-bearing nanospinels identified in the matrix of primitive carbonaceous chondrites have elevated $^{54}$Cr/$^{52}$Cr ratios of up to 50x the solar composition (Qin et al., 2010; Dauphas et al., 2010; Nittler et al., 2018). To assess whether these grains can account for the observed $\varepsilon^{54}$Cr variations among chondrules, we calculated the number of nanospinel grains that would need to be added to an average CV chondrule to generate the observed 1-2 $\varepsilon^{54}$Cr variations. Using typical values for CV chondrules (1 mg; 3000 µg/g Cr) and presolar nanospinels (80-200 nm; 2.5x solar $^{54}$Cr/$^{52}$Cr; Dauphas et al., 2010; Nittler et al., 2018), we find that tens of thousands of grains would be needed to generate the observed 1-2 $\varepsilon^{54}$Cr variations among the chondrules. The basic picture does not change even when grains with the highest reported anomalies (56x solar) are used in the calculation, since the $^{54}$Cr anomalies of the nanospinel grains are inversely correlated with their size (Nittler et al. 2018). Therefore, if presolar nanospinels were the only source of $\varepsilon^{54}$Cr variations among chondrules, then the amount of presolar grains mixed into different chondrules would have to vary by tens of thousands of grains from one chondrule to the next. Such a scenario is highly unrealistic in terms of mixing dynamics of dust in the solar nebula, and contrasts with the overall rarity of presolar grains in the meteorite record. As such, presolar nanospinels do not seem to be responsible for the $\varepsilon^{54}$Cr variability among individual chondrules.

As was shown by Gerber et al. (2017), Ti isotope variations in CC chondrules were most likely caused by incorporation of variable amounts of refractory inclusions such as CAIs. However, although CAIs have well resolved $\varepsilon^{54}$Cr excesses, they contain too little Cr to significantly affect the Cr isotopic composition of chondrules, and so admixture of CAIs to chondrule precursors results in a strongly curved mixing hyperbola in $\varepsilon^{50}$Ti vs. $\varepsilon^{54}$Cr space with little overall variations in $\varepsilon^{54}$Cr (Fig. 4d). Accordingly, the chondrules of this study do not plot on a mixing line between NC chondrules and CAIs (Fig. 4d-f). Moreover, there are chondrules with $\varepsilon^{54}$Cr excesses but no $\varepsilon^{50}$Ti anomalies, indicating that the $\varepsilon^{54}$Cr excesses are unrelated to the addition of CAIs. These observations rule out CAI admixture as a significant driver for $\varepsilon^{54}$Cr variations among chondrules, and shows that the $^{50}$Ti and $^{54}$Cr anomalies are carried by distinct materials.

Amoeboid olivine aggregates (AOAs) are another component that is abundant in carbonaceous chondrites and may be responsible for the $\varepsilon^{54}$Cr variations among chondrules. Of note, AOAs are characterized by similar $\varepsilon^{54}$Cr and $\varepsilon^{50}$Ti excesses than CAIs (Trinquier et al., 2009), but can have much higher Cr/Ti (Komatsu et al., 2001; Sugiura et al., 2009), reflecting their less refractory character compared to CAIs. Moreover, there is growing evidence that CC chondrules inherited primitive $^{16}$O-rich precursors including not only CAIs but also AOA-like relict silicate grains, suggesting admixture of non-CAI but isotopically anomalous material to chondrule precursors (Kita et al., 2010; Marrocchi et al., 2019, 2018a). This is also consistent with the observation that samples of this study include chondrules containing relict olivine grains with negative $\Delta^{17}$O of up to -14 ‰. Moreover, when the Cr, Ti, and O isotope variations among the chondrules are considered together (Fig. 4d-f), it is evident that addition of CAIs alone cannot account for the observed variations. Instead, components with CAI-like isotopic compositions but less refractory chemical compositions, such as AOAs, must have been part of chondrule precursors.

To more quantitatively assess as to whether the addition of AOA-like material is consistent with the observed isotopic variability among chondrules, we calculated mixing lines between two chemically and isotopically distinct dust components. One component is characterized by a chondritic chemical and NC-like isotopic composition. The other component has a CAI/AOA-like isotopic composition and is characterized by variable refractory to non-refractory chemical compositions; this range in compositions is reflected in increasing Cr/Ti from typical CAI-like values to the more elevated values observed for AOAs. The results of the mixing calculations show that heterogeneous mixing of different chemical components from two bulk dust reservoirs reproduces the entire range of observed chondrule compositions quite well (Fig. 4d-f). For instance, while the extreme composition of the Al-rich chondrule MSc14 from Allende ($\varepsilon^{50}$Ti of ~7, $\varepsilon^{54}$Cr ~0.4) reflects admixture of refractory, CAI-like material, chondrules with $\varepsilon^{54}$Cr excesses but NC-like $\varepsilon^{50}$Ti (e.g., Allende chondrules MSc 3 and MSc 8) are consistent with addition of material with AOA-like Cr/Ti (Fig. 4d). The mixing calculations involving O isotopes are less straight forward because the O concentrations of the mixing endmembers are not well known. Nevertheless, assuming a reasonable range of O/Cr and O/Ti ratios for the mixing endmembers [O concentrations are based on the typical modal mineralogy of OCs (McSween et al., 1991), CAIs (Grossmann, 1975) and AOAs (assuming pure forsterite)], we find that all CC chondrules fall within the range of possible mixing trajectories (Fig. 4e). Finally, it is noteworthy that although the NC chondrules of this study fully plot within the compositional fields of bulk NC bodies, they nevertheless seem to define a weak trend towards the isotopic composition of CAIs/AOAs (Fig. 4). This may suggest that small amounts of isotopically CAI-like material were also present among the precursors of NC chondrules. However, investigating the relevance and origin of this trend will require Ti and Cr isotope measurements of rare OC chondrules with relict $^{16}$O-rich grains (Kita et al. 2010).

In summary, the Cr-Ti-O isotopic heterogeneity among single chondrules can be accounted for by mixing of an NC dust component with chemically variable dust components having a CAI/AOA-like isotopic composition. This mixing model is consistent with the higher abundance of CAIs and AOAs in carbonaceous chondrites, and the rarity of these components in NC chondrites.

*4.2 Chondrule transport versus precursor heterogeneity*

Because the range of $\varepsilon^{54}$Cr values observed among CV chondrules is similar to the range measured among bulk meteorites, Olsen et al. (2016) argued that a chondrule's $\varepsilon^{54}$Cr reflects its specific formation location within the disk. This interpretation implies that chondrules with distinct $\varepsilon^{54}$Cr formed in different regions of the disk. As such, the variable $\varepsilon^{54}$Cr of CV chondrules would in turn require formation of chondrules from a given chondrite group across large areas of the disk, and the subsequent disk-wide transport of the chondrules to the accretion region of their respective host chondrites. This interpretation of the chondrules' isotopic compositions as source signatures is very different from the interpretation put forward above, in which the isotopic heterogeneities are inherited signatures of chondrule precursors. As noted above, while it is difficult to distinguish between these two interpretations based solely on isotopic data for a single element, the combined Cr-Ti-O isotopic data from this study allow distinguishing between source and precursor signatures.

If the isotopic variations would reflect distinct source regions of chondrules, then their $\Delta^{17}$O, $\varepsilon^{50}$Ti, and $\varepsilon^{54}$Cr signatures should be correlated. For example, a chondrule from the inner Solar System (i.e., the NC reservoir) that was transported to the accretion region of CV chondrites should not only exhibit an NC-like $\varepsilon^{54}$Cr, but also NC-like $\Delta^{17}$O and $\varepsilon^{50}$Ti. Conversely, a CC chondrule from the outer Solar System (i.e., the CC reservoir) should have CC-like $\Delta^{17}$O, $\varepsilon^{50}$Ti, and $\varepsilon^{54}$Cr anomalies. However, among the CV chondrules there are samples with NC-like $\varepsilon^{50}$Ti and CC-like $\varepsilon^{54}$Cr, and there are also CV chondrules with CC-like $\varepsilon^{50}$Ti but NC-like $\varepsilon^{54}$Cr. Moreover, in $\Delta^{17}$O versus $\varepsilon^{54}$Cr space, all CC chondrules plot close to the field of bulk CC chondrites (Fig. 4b,e). Thus, although CV chondrules with NC-like $\varepsilon^{54}$Cr exist, the combined $\Delta^{17}$O–$\varepsilon^{54}$Cr of these chondrules nevertheless places their origin in the CC reservoir. Combined these observations highlight that the Cr-Ti-O isotopic variability among CV chondrules cannot be tied to an origin from distinct disk regions in a self-consistent way. These data are, therefore, inconsistent with variable formation regions of CV (and other CC) chondrules and subsequent transport of chondrules through the disk. By contrast, the Cr-Ti-O isotopic variations can readily be attributed to local heterogeneities among chondrule precursors, resulting from the incorporation of chemically diverse and isotopically CAI-like material into the dust aggregates that became chondrules. This interpretation is also consistent with mixing models that employ bulk chemical compositions of chondrules, which also favor local precursor heterogeneity over chondrule transport (Hezel and Parteli, 2018).

***4.3 Relation between isotopic anomalies in chondrules and the NC-CC dichotomy of the disk***
18

Although the isotope anomalies in chondrules do not trace spatially distinct origins and large-scale transport of the chondrules themselves, they nevertheless provide evidence for transport and mixing among their precursors. This is because the degree of isotopic variability among chondrules of a given chondrite group reflects the amount of isotopically anomalous precursor materials present in a given region of the disk, and how well this material was mixed at the chondrule precursor scale. The isotopic heterogeneities among chondrules, therefore, provide information on the spatial and temporal evolution of the solar protoplanetary disk.

Recently, it has been argued that the isotopic dichotomy between the NC and CC reservoirs of the accretion disk results from a change in the isotopic composition of infalling material from the Sun's parental molecular cloud (Burkhardt et al., 2019; Nanne et al., 2019). In this model, early infalling material was characterized by an isotopic composition as measured in CAIs, whereas the isotopic composition of the later infall had an NC-like isotopic composition. The outward transport of early infalling material by viscous spreading (Hueso and Guillot, 2005; Pignatale et al., 2018; Yang and Ciesla, 2012) resulted in an early disk having a CAI-like isotopic composition (termed IC for Inclusion-like Chondritic reservoir; Burkhardt et al., 2019). Later infall onto the disk, now with an NC-like isotopic composition then changed the composition of the inner disk, while the outer disk retained a memory of the isotopic composition of the early disk. Ultimately, these processes led to the formation of the CC reservoir, representing the outer disk whose composition can be regarded as a mixture of IC and NC material, and the NC reservoir, representing the inner disk, whose composition is dominated by the later infall.

The combined Cr-Ti-O isotopic data for chondrules are consistent with this model, because they reveal limited variability in NC chondrules and variable $\delta^{16}O$ (i.e., low $\Delta^{17}O$), $\varepsilon^{54}Cr$, and $\varepsilon^{50}Ti$ enrichments in CC chondrules. Since chondrules provide a snapshot of the dust composition at their formation location (Scott, 2007; Budde et al., 2016b; Bollard et al., 2017;

Morris and Boley, 2018), the relatively homogeneous isotopic composition of NC chondrules indicates that they formed in a well-mixed reservoir dominated by NC dust. This is consistent with the idea that the inner disk predominantly consists of material from the later infall (Burkhardt et al., 2019; Nanne et al., 2019), and with the rarity of CAIs and CAI-like refractory material in NC chondrites (Ebert et al., 2018). By contrast, CC chondrules formed in a region of the disk which contains material from both the early and late infall. The data of this study indicate that this mixture of early and late infalling material was heterogeneous at the level of individual CV chondrules, which did not only incorporate various amounts of CAIs, but also less refractory material with CAI-like isotopic compositions. This is consistent with the higher abundance of CAIs and AOAs in CC over NC meteorites, and with the idea that these components represent remnants of the IC reservoir, which had formed during the early infall and contained not only CAIs, but also less refractory material (Burkhardt et al., 2019; Nanne et al., 2019). Finally, within the CC reservoir, the isotopic variations among chondrules seem to show a progressive homogenization from CV and CO to CR chondrules (Gerber et al., 2017; Zhu et al., 2019). This is consistent with the late formation times of ~3.7 Ma after CAIs for CR chondrules compared to ~2.2 Ma after CAIs for CV chondrules (Schrader et al., 2017; Nagashima and Komatsu, 2017; Budde et al., 2016; Budde et al., 2018), and may indicate the progressive homogenization of IC and NC material within the outer disk.

*4.4 Origin of isotope anomalies in bulk carbonaceous chondrites*

The evidence for mixing between IC and NC components in the outer disk has implications for understanding the generation of isotope anomalies among bulk carbonaceous chondrites. In the $\varepsilon^{50}$Ti–$\Delta^{17}$O and $\varepsilon^{50}$Ti–$\varepsilon^{54}$Cr diagrams (Fig. 4), CC chondrules plot within and outside the fields of bulk CC chondrites, reflecting the strong control of CAIs on particular $\varepsilon^{50}$Ti (Gerber et al., 2017). Variable CAI abundances also account for much of the $\varepsilon^{50}$Ti variations among bulk carbonaceous chondrites (Trinquier et al., 2009). By contrast, in $\Delta^{17}$O–$\varepsilon^{54}$Cr space, CC chondrules plot close to the compositional field of bulk CC chondrites. As such, the chondrule data can help assess the origin of $\varepsilon^{54}$Cr variations among bulk CCs.

Trinquier et al. (2007) observed a correlation between $\Delta^{17}$O and $\varepsilon^{54}$Cr among bulk carbonaceous chondrites (see Fig. 4) and argued that this correlation reflects mixing between two isotopically distinct reservoirs. However, the origin of these reservoirs, and whether they can be tied to specific meteorite components, remained ambiguous. More recently, it was argued that the $\Delta^{17}$O–$\varepsilon^{54}$Cr correlation results from mixing CI-like dust with a hypothetical $^{16}$O-rich (i.e., CAI-like $\Delta^{17}$O) but $^{54}$Cr-poor ($\varepsilon^{54}$Cr ≈ –10) component (Alexander et al. 2019; Zhu et al. 2019). Direct evidence for the existence of such a $^{54}$Cr-poor component is lacking, however. None of the CC chondrules of this study plot outside the range of mixing trajectories between the NC and IC reservoirs, and so these data also provide no evidence for the existence of material with strongly negative $\varepsilon^{54}$Cr. However, mixing between NC and IC materials does not easily account for the $\Delta^{17}$O–$\varepsilon^{54}$Cr array of bulk carbonaceous chondrites, which is almost perpendicular to the NC-IC mixing trajectories (Fig. 5). Either this array is inherited from the NC reservoir, where bulk meteorites plot along a similar, parallel trend, or the array reflects variable mixing between CI-like material, characterized by the highest $\varepsilon^{54}$Cr and $\Delta^{17}$O, and chondrules with lower $\varepsilon^{54}$Cr and $\Delta^{17}$O. Of note, for pooled Allende chondrule separates, each consisting of hundreds of chondrules, we obtained $\varepsilon^{54}$Cr = 0.58±0.11 (Table 1). Combined with the average $\Delta^{17}$O = –5.4±0.9 of this study, this shows that CV chondrules plot at the lower left of the $\varepsilon^{54}$Cr–$\Delta^{17}$O array defined by bulk CCs. As such, mixing between CI-like material and a common "chondrule-component" may account for the correlated $\varepsilon^{54}$Cr–$\Delta^{17}$O variations among bulk carbonaceous chondrites. Such a mixing scenario would be consistent with chemical variations among CC chondrites (e.g., Braukmüller et al., 2018; Alexander, 2019), and also with the correlation of mass-dependent Te isotope compositions and $\varepsilon^{54}$Cr anomalies observed among bulk carbonaceous chondrites (Hellmann et al., 2020).

## 5. Conclusions

The first combined Cr-Ti-O isotope data set for individual chondrules reveals a fundamental difference between chondrules from non-carbonaceous (NC) and carbonaceous chondrites (CC). Whereas NC chondrules display relatively homogeneous $\Delta^{17}O$, $\varepsilon^{50}Ti$, and $\varepsilon^{54}Cr$, which are indistinguishable from the compositions of their respective host chondrites, CC chondrules have more variable $\varepsilon^{50}Ti$, $\varepsilon^{54}Cr$, and $\Delta^{17}O$. Importantly, although the compositions of CC and NC chondrules may overlap for either $\varepsilon^{50}Ti$, $\varepsilon^{54}Cr$, or $\Delta^{17}O$, in multi-isotope space none of the measured CC chondrules plots in the compositional field of NC chondrites, and none of the NC chondrules plots within the field of CC chondrites. These data are inconsistent with a prior proposal for the disk-wide transport of chondrules across the NC and CC reservoirs, and indicate that there has been no significant exchange of chondrules between these reservoirs. Instead the entire range of isotopic anomalies in chondrules is attributable to the heterogeneous nature of their precursors, and can be accounted for by mixing between NC-like dust and a chemically diverse dust component that was isotopically similar to CAIs and AOAs. The same mixing processes, but on a larger, disk-wide scale, were likely responsible for establishing the distinct isotopic compositions of the NC and CC reservoirs, which represent in inner and outer disk, respectively.

**Acknowledgments**

We are grateful to NASA and the Naturhistorisches Museum Wien for providing samples for this study, and thank W. McKinnon for editorial handling and the reviewers for their constructive comments. Acquisition of the TIMS was made possible through a Sofja Kovalevskaja award from Alexander von Humboldt Foundation (G.A.B.), which is gratefully acknowledged. Some parts of this study were performed under the auspices of the US DOE by Lawrence Livermore National Laboratory under Contract DE-AC52-07NA27344 with release number LLNL-JRNL-802021. Funded by the Deutsche Forschungsgemeinschaft (DFG, German Research Foundation) – Project-ID 263649064 – TRR 170". This is TRR pub. no. 109.

**Table 1:** Chromium, titanium and oxygen isotopic and concentration data for terrestrial samples, bulk chondrites and chondrules.

| Sample | Type[a] | Ø [mm][b] | Weight [mg][c] | Cr [ppm][d] | n[e] | $\varepsilon^{53}$Cr ± 2σ[f] | $\varepsilon^{54}$Cr ± 2σ[f] | Ti [ppm][b] | $\varepsilon^{50}$Ti[f] ± 2σ | $\Delta^{17}$O ± 2σ |
|---|---|---|---|---|---|---|---|---|---|---|
| **Terrestrial** | | | | | | | | | | |
| DTS2b | Silicates[g] | | | 27 | 1551 | 8 | 0.12 ± 0.07 | 0.16 ± 0.13 | | |
| DTS2b | Chromites[g] | | | | | 7 | -0.03 ± 0.20 | -0.02 ± 0.34 | | |
| JA-2 | | | | 161 | 423 | 5 | 0.03 ± 0.16 | 0.13 ± 0.28 | | |
| BIR1a | | | | 240 | 392 | 3 | 0.08 ± 0.17 | 0.07 ± 0.30 | | |
| Average | | | | | | | **0.05 ± 0.13** | **0.09 ± 0.16** | | |
| **OC (Ragland - LL3.4)** | | | | | | | | | | |
| MSc42 | Type II, POP | 2.7 | 17 | 4108 | 1 | 0.40 ± 0.10 | -0.24 ± 0.24 | 601 | -0.43 ± 0.14 | 1.28 ± 1.24 |
| MSc43 | Type II, RP | 2.1 | 14 | 4613 | 5 | 0.45 ± 0.17 | -0.11 ± 0.18 | 327 | -0.41 ± 0.13 | 0.82 ± 0.63 |
| MSc44 | Type II, RP | 2.4 | 13 | 4693 | 3 | 0.25 ± 0.15 | -0.34 ± 0.31 | 550 | -0.48 ± 0.13 | 1.58 ± 2.25 |
| MSc45 | Type II, RP(O) | 2.6 | 18 | 3959 | 4 | 0.37 ± 0.16 | -0.14 ± 0.34 | 803 | -1.00 ± 0.12 | 1.83 ± 0.97 |
| MSc49 | Type II, RP | 1.4 | 1.8 | 4316 | 3 | 0.74 ± 0.13 | -0.39 ± 0.32 | 439 | -0.67 ± 0.56 | 1.43 ± 0.60 |
| MSc50 | Type II, PO | 1.4 | 2.4 | 3870 | | | | 741 | -0.53 ± 0.08 | 1.15 ± 0.39 |
| MSc51 | Type II, C | 1.0 | 1.2 | 4593 | 2 | 0.47 ± 0.13 | -0.51 ± 0.32 | 553 | -0.58 ± 0.08 | 1.79 ± 0.83 |
| MSc52 | Type II, POP | 1.4 | 2.8 | | | | | 800 | -0.60 ± 0.09 | 1.09 ± 1.29 |
| MSc53 | Type II, RP | 1.5 | 1.2 | | | | | 472 | -0.24 ± 0.15 | 1.07 ± 1.09 |
| Average | | **1.8** | | **4307** | | **0.45 ± 0.33** | **-0.29 ± 0.31** | **587** | **-0.55 ± 0.42** | **1.34 ± 0.69** |
| | | | | | | | | | | |
| Std8 | Whole rock | | | 51 | 3034 | 4 | 0.21 ± 0.06 | -0.31 ± 0.19 | 606 | -0.70 ± 0.13 |
| **EC (MAC02837 - EL3)** | | | | | | | | | | |
| MSc55 | Type I, PP | 1.3 | 1.3 | 2200 | | | | 443 | -0.13 ± 0.11 | -0.20 ± 0.54 |
| MSc57 | Type I, PP | 1.2 | 0.9 | 2217 | | | | 546 | -0.21 ± 0.13 | -0.02 ± 0.44 |
| MSc58 | Type I, RP | 0.9 | 0.7 | 1865 | | | | 408 | -1.47 ± 0.19 | -0.08 ± 0.81 |
| MSc61 | Type I, PP | 1.7 | 4.4 | 796 | 1 | 0.18 ± 0.06 | 0.01 ± 0.19 | 137 | -0.28 ± 0.08 | 0.38 ± 0.94 |
| Average | | **1.3** | | **1769** | | | | **384** | **-0.52 ± 1.27** | **0.02 ± 0.51** |
| | | | | | | | | | | |
| Std9 | Whole rock | | | 54 | 2843 | 1 | 0.16 ± 0.06 | 0.02 ± 0.19 | 486 | -0.40 ± 0.12 |
| **CV (Allende - CV3)** | | | | | | | | | | |
| MSc2 | Type I, PO | 1.5 | 1.5 | 2493 | 5 | 0.12 ± 0.25 | 1.24 ± 0.14 | 1190 | 4.85 ± 0.16 | -5.21 ± 1.29 |
| MSc3 | Type I, PO(P) | 1.2 | 1.7 | 2670 | 5 | 0.07 ± 0.16 | 0.71 ± 0.19 | 793 | -0.15 ± 0.14 | -5.95 ± 0.75 |
| MSc5 | Type I, BO | 1.2 | 1.3 | 3295 | 4 | 0.05 ± 0.17 | 1.16 ± 0.21 | 1401 | 1.73 ± 0.18 | -4.82 ± 1.22 |
| MSc8 | Type I, POP | 1.9 | 5.7 | 3121 | 7 | 0.13 ± 0.10 | 1.43 ± 0.16 | 775 | -0.05 ± 0.08 | -2.98 ± 1.21 |
| MSc10 | Type I, PO | 1.4 | 2.0 | 2844 | 6 | 0.07 ± 0.12 | 0.82 ± 0.15 | 1015 | 1.30 ± 0.11 | -4.99 ± 1.90 |
| MSc12 | Type I, PO | 1.8 | 2.9 | 3307 | 6 | 0.07 ± 0.12 | 0.09 ± 0.11 | 953 | 1.56 ± 0.20 | -5.43 ± 1.45 |
| MSc14 | Al-rich | 1.8 | 4.2 | 2987 | 3 | -0.10 ± 0.13 | 0.44 ± 0.32 | 1760 | 7.32 ± 0.09 | -5.74 ± 1.13 |
| Average | | **1.5** | | **2960** | | **0.06 ± 0.15** | **0.84 ± 0.95** | **1127** | **2.36 ± 5.48** | **-5.02 ± 1.97** |
| | | | | | | | | | | |
| AL 25 | Chondr. fraction | 0.25-1.6 | | 302 | | 17 | 0.07 ± 0.09 | 0.55 ± 0.23 | 1204 | 2.50 ± 0.12 |
| AL 32 | Chondr. fraction | 0.5-1.0 | | 384 | | 10 | 0.03 ± 0.08 | 0.65 ± 0.19 | 1161 | 2.17 ± 0.14 |
| AL 36 | Chondr. fraction | 0.5-1.0 | | 518 | | 15 | 0.05 ± 0.08 | 0.55 ± 0.18 | 1029 | 2.24 ± 0.15 |
| Average | | | | | | | **0.05 ± 0.04** | **0.58 ± 0.11** | **1131** | **2.30 ± 0.35** |
| | | | | | | | | | | |
| AL 26 | Matrix fraction | | | 506 | | 14 | 0.30 ± 0.07 | 1.18 ± 0.14 | 656 | 2.68 ± 0.07 |
| AL 27 | Matrix fraction | | | 394 | | 17 | 0.15 ± 0.04 | 0.98 ± 0.13 | 663 | 1.91 ± 0.19 |
| AL 42 | Matrix fraction | | | 469 | | 8 | 0.17 ± 0.09 | 1.01 ± 0.19 | 696 | 2.25 ± 0.15 |
| Average | | | | | | | **0.21 ± 0.16** | **1.06 ± 0.22** | **672** | **2.28 ± 0.78** |
| | | | | | | | | | | |
| Allende 2 | Whole rock | | | 37 | | 6 | 0.12 ± 0.12 | 0.96 ± 0.22 | | |
| Allende 2 | Whole rock | | | 41 | | 4 | 0.11 ± 0.10 | 1.06 ± 0.30 | | |
| Ti1 | Whole rock | | | 37 | 3970 | 4 | 0.13 ± 0.20 | 1.06 ± 0.34 | 731 | 3.31 ± 0.14 |
| Ti2 | Whole rock | | | 32 | 4013 | 12 | 0.20 ± 0.06 | 0.99 ± 0.11 | 826 | 3.22 ± 0.34 |
| Std1 | Whole rock | | | 10 | 3198 | 6 | 0.19 ± 0.07 | 1.18 ± 0.18 | 729 | 3.20 ± 0.16 |
| Average | | | | | **3727** | | **0.15 ± 0.08** | **1.05 ± 0.17** | **762** | **3.24 ± 0.12** |
| **CR (GRA06100 - CR2)** | | | | | | | | | | |
| MSc72 | Type I, POP | 0.9 | 0.9 | 3984 | 2 | 0.06 ± 0.13 | 1.25 ± 0.30 | 873 | 2.21 ± 0.16 | -2.28 ± 0.60 |
| MSc73 | Type I, POP | 0.7 | 0.4 | 4952 | | | | 842 | 2.13 ± 0.26 | -2.58 ± 1.69 |
| MSc74 | Type I, POP | 1.1 | 0.9 | 3880 | 2 | 0.27 ± 0.13 | 1.39 ± 0.30 | 624 | 1.98 ± 0.23 | -1.72 ± 0.47 |
| MSc75 | Type I, POP | 1.1 | 1.0 | 5617 | | | | 977 | 1.79 ± 0.16 | -3.65 ± 0.72 |
| MSc76 | Type I, POP | 1.2 | 1.9 | 3778 | 2 | 0.21 ± 0.13 | 1.20 ± 0.30 | 782 | 1.41 ± 0.15 | -3.88 ± 0.86 |
| MSc77 | Type I, POP | 1.8 | 4.6 | 3706 | 2 | 0.24 ± 0.13 | 1.37 ± 0.30 | 789 | 1.21 ± 0.08 | -0.89 ± 1.00 |
| Average | | **1.1** | | **4320** | | **0.20 ± 0.19** | **1.30 ± 0.18** | **815** | **1.79 ± 0.80** | **-2.50 ± 1.14** |
| | | | | | | | | | | |
| Std10 | Whole rock | | | 51 | 3096 | 3 | 0.26 ± 0.13 | 1.23 ± 0.30 | 715 | 3.26 ± 0.09 |
| **CR (NWA 801 - CR2)** | | | | | | | | | | |
| MSc78 | Type I, POP | 1.6 | 1 | 3393 | | | | 642 | 3.39 ± 0.17 | -1.75 ± 1.10 |

[a] Type I chondrules have Fa-content <10, type II chondrules >10; PP—porphyritic pyroxene, PO—porphyritic olivine, POP—porphyritic olivine-pyroxene, BO—barred olivine, RP—radial pyroxene, C—cryptocristalline
[b] data from Gerber et al. (2017)
[c] mass denotes digested sample mass, for chondrules this represents between 80 and 95% of chondrule mass
[d] error on Cr concentration measurements is estimated to be 10%
[e] number of individual Cr isotope measurements
[f] chondrule data: for n≥4 the 95% confidence interval is used, for n≤4 the 2s.d. of the NIST3112a standard of the respective session is used, for averages the 2s.d. is given
[g] bulk powders of DTS2b were digested in aqua regia and subsequent HF-HNO$_3$. The residual chromites were digested in Parr bombs with 2ml conc. HNO$_3$ at 190°C for 4 days

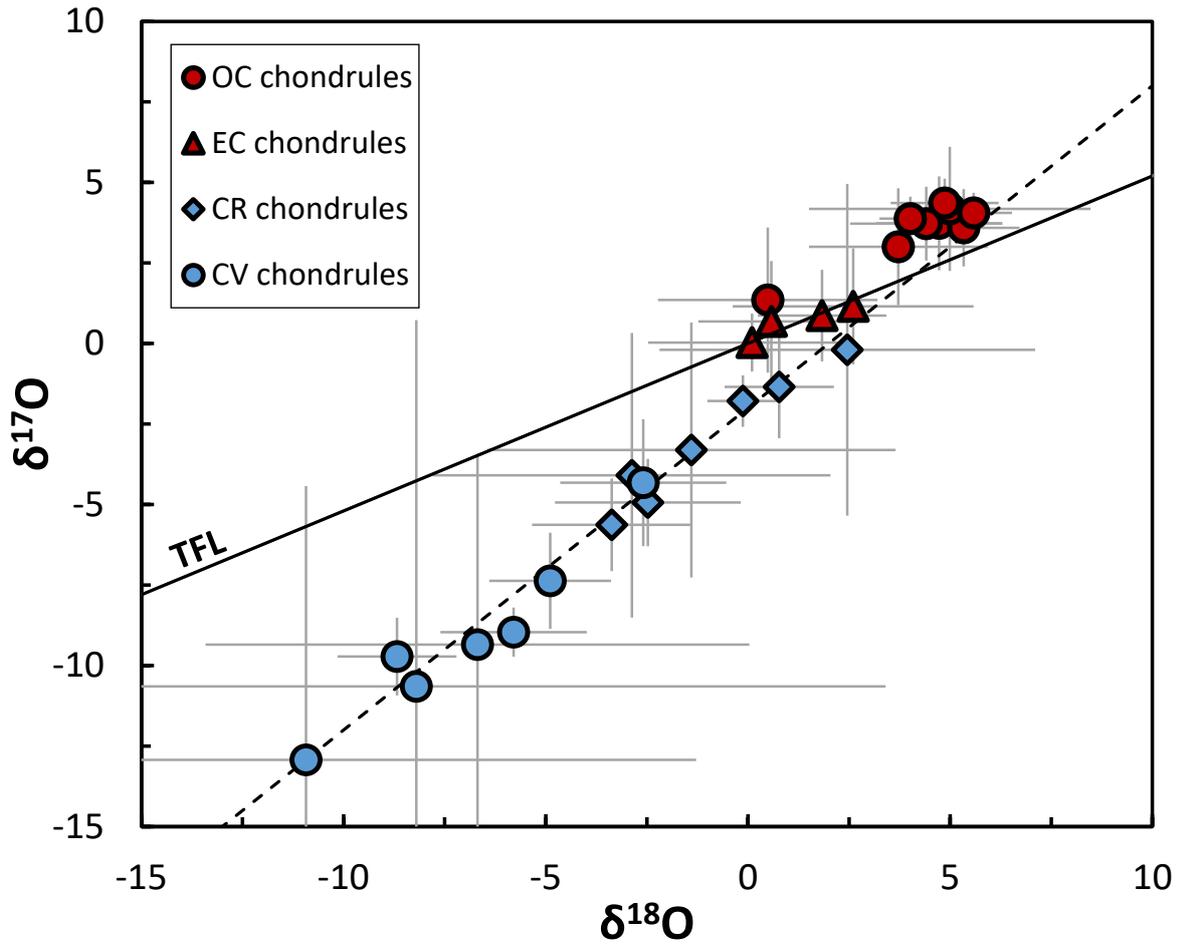

**Figure 1:** Oxygen three-isotope plot. CC chondrules plot along the CCAM line with a slope of approximately 1. The comparatively large errors are due to intra-chondrule heterogeneity with single olivine grains being isotopically distinct from one another. OC and EC chondrules do not display significant intra-chondrule variation and plot on or above the terrestrial fractionation line (TFL) with a slope of 0.52.

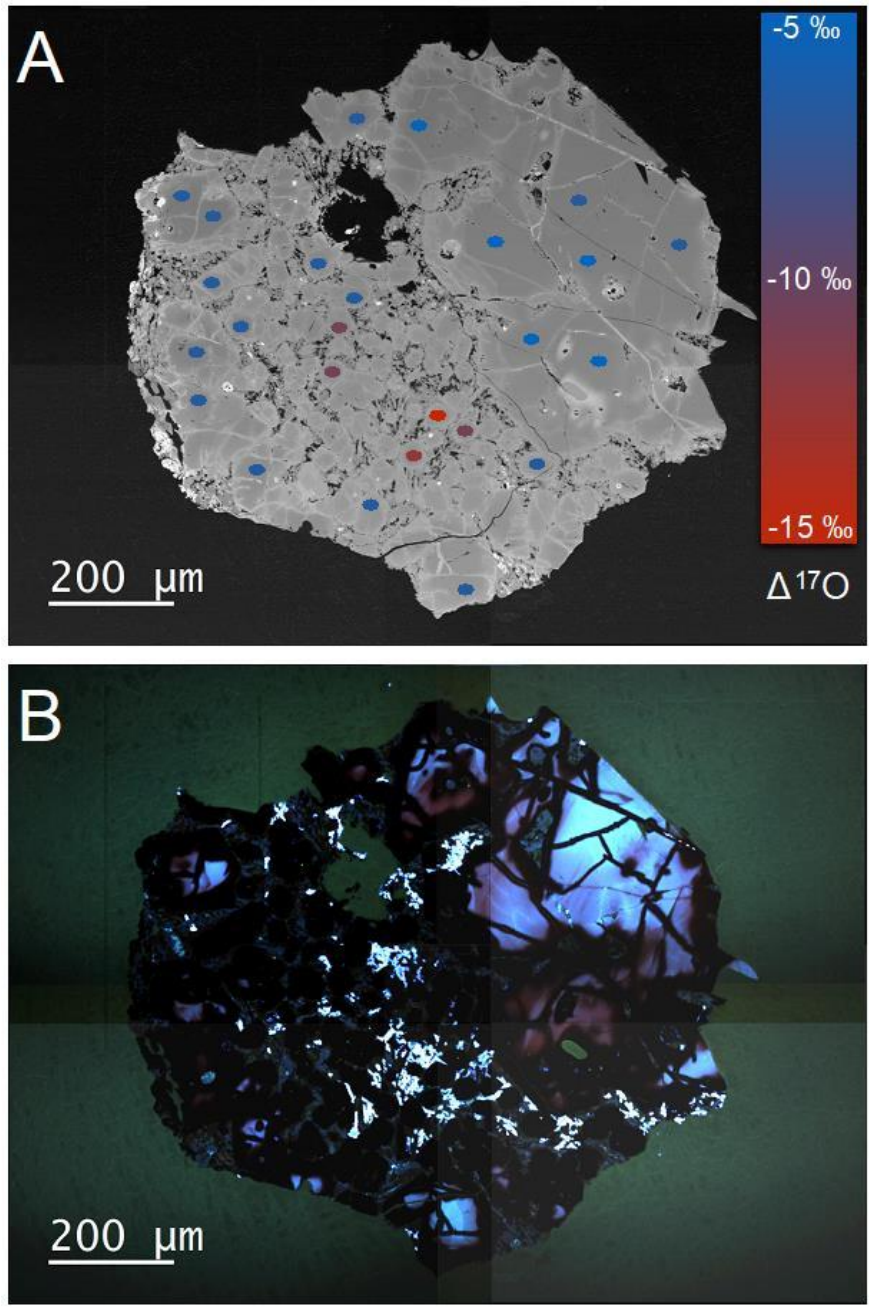

**Figure 2:** (a) Back-scattered electron image of PO chondrule MSc10 from Allende (CV3) showing the locations of ion probe measurements (colored ellipses). The colors indicate the range of $\Delta^{17}O$ values. (b) CL image revealing internal structures of olivine grains. $^{16}O$-rich relict olivine grains show no CL emission due to low abundances of CL activators (e.g., Ca, Al, Ti) while host olivine grains show CL emission induced by higher concentration of Ca, Al, and Ti.

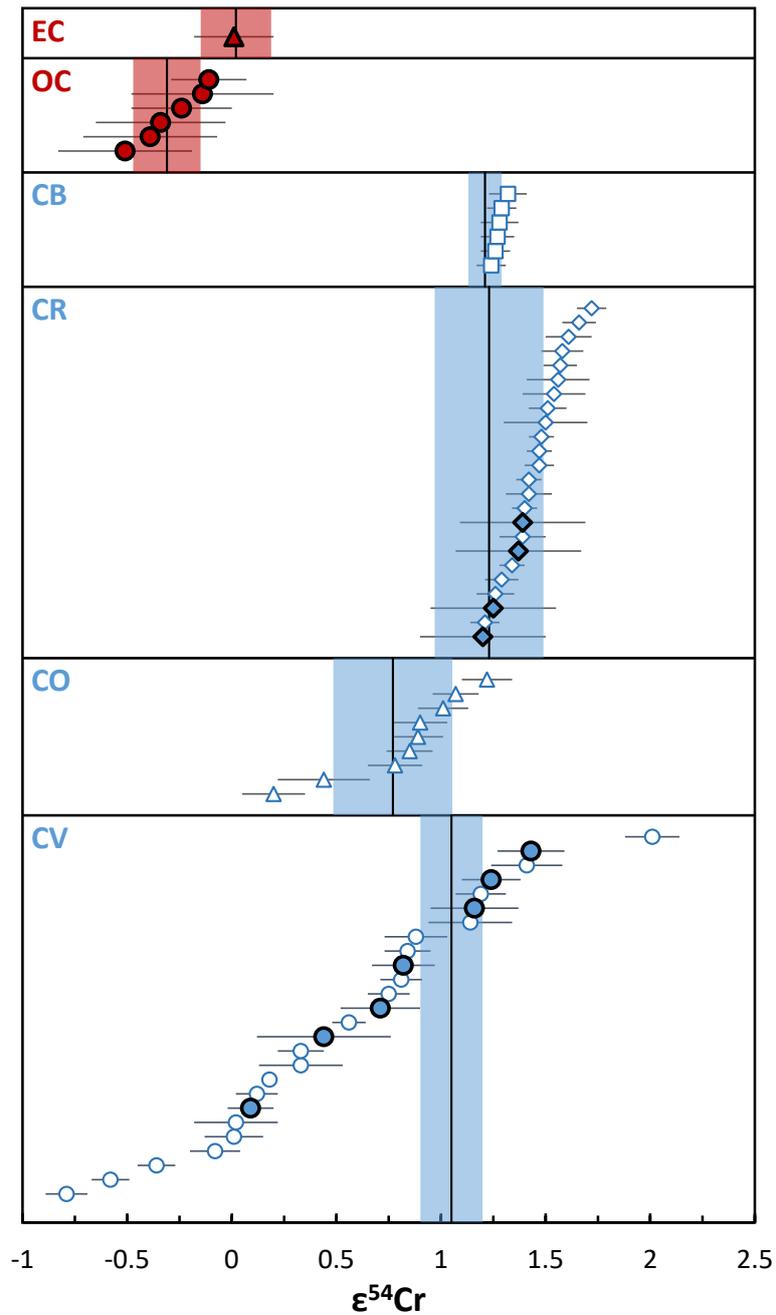

**Figure 3:** $\varepsilon^{54}$Cr values of individual chondrules. Compositions of respective bulk chondrites are shown as solid lines; shaded areas around these lines indicate 2 s.d. uncertainties of the bulk values. Bold symbols represent data from this study, open symbols are literature data (Trinquier et al., 2008; Yamashita et al., 2010; Connelly et al., 2012; Olsen et al., 2016; Zhu et al., 2019). Bulk chondrite values are from this study (EC, OC, CR, and CV), and Qin et al. (2010) (CB, CO).

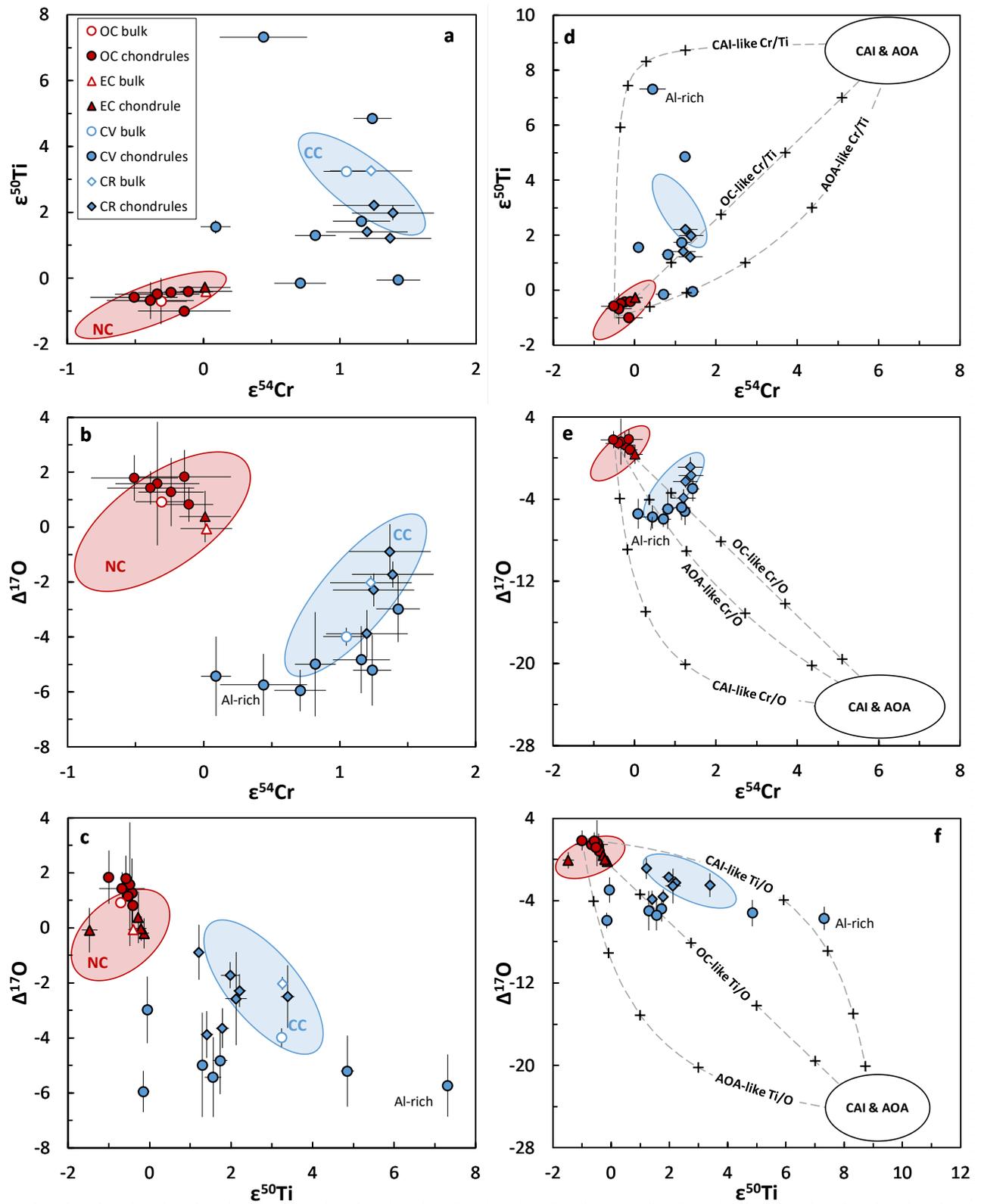

**Figure 4:** $\Delta^{17}O$, $\varepsilon^{50}Ti$, and $\varepsilon^{54}Cr$ of individual chondrules in comparison to bulk NC and CC meteorites (a-c) and CAIs/AOAs (d-f). The NC (red) and CC (blue) fields represent the range of anomalies measured for bulk meteorites from the NC and CC reservoir, respectively. The NC and CC fields as well as representative CAI isotopic compositions are based on data compiled in Burkhardt et al. (2019). Bulk chondrite data points (open symbols) are only shown for samples from which individual chondrules (filled symbols) were analyzed in this study. (a)-(c) NC chondrules display relatively homogeneous $\Delta^{17}$O, $\varepsilon^{50}$Ti, and $\varepsilon^{54}$Cr, which are indistinguishable from the compositions of their respective host chondrites, while CC chondrules have more variable $\varepsilon^{50}$Ti, $\varepsilon^{54}$Cr, and $\Delta^{17}$O. Note that although the compositions of CC and NC chondrules may overlap for either $\varepsilon^{50}$Ti, $\varepsilon^{54}$Cr, or $\Delta^{17}$O, in multi-isotope space no CC chondrule plots within the NC field, and no NC chondrule plots within the CC field. Also note that most CC meteorites are affected by aqueous alteration which may modify their pristine $\Delta^{17}$O through the interaction with $^{16}$O poor water(-ice) (Piralla et al., 2020). Alteration corrected $\Delta^{17}$O values for bulk carbonaceous chondrites would thus plot at slightly lower $\Delta^{17}$O. (d)-(f) The composition of all chondrules can be reproduced by mixing between an NC component with chondritic Cr/Ti/O ratios and an isotopically CAI/AOA-like component (IC) having more variable Cr/Ti/O ratios. Crosses on mixing lines indicate 20% percent steps of IC component in NC-IC mixture.

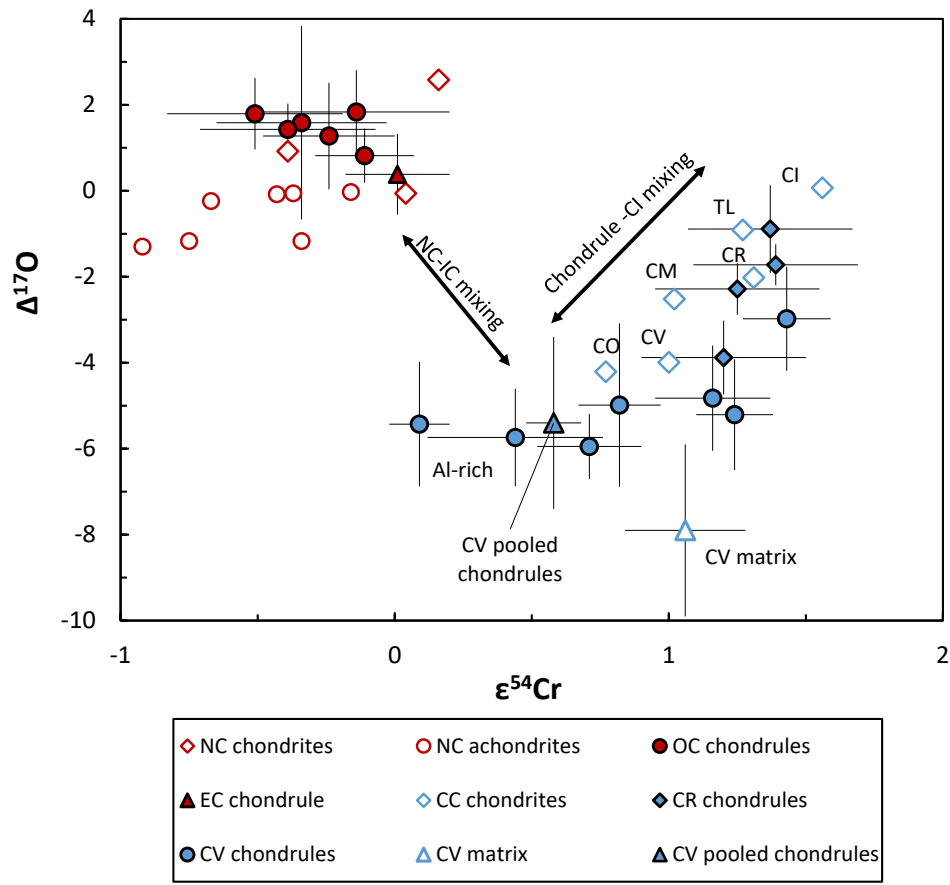

**Figure 5:** $\Delta^{17}O$ vs. $\varepsilon^{54}Cr$ of chondrules and bulk meteorites. NC (red) and CC (blue) bodies (open symbols) define distinct fields. While the offset between NC and CC meteorites most likely reflects mixing between NC and IC materials, the broad correlation of $\Delta^{17}O$ and $\varepsilon^{54}Cr$ among CC materials most likely reflects mixing between average CC chondrules and CI-like dust. Bulk meteorite data from compilation in Burkhardt et al. (2019). For clarity error bars on bulk meteorite data were omitted. Note that most CC meteorites are affected by aqueous alteration which may modify their pristine $\Delta^{17}O$ through the interaction with $^{16}O$ poor water(-ice) (Piralla et al., 2020). Alteration corrected $\Delta^{17}O$ values for bulk carbonaceous chondrites would thus plot at slightly lower $\Delta^{17}O$. Data for $\Delta^{17}O$ of CV matrix was taken from Piralla et al. (2020). TL: Tagish Lake.

746 **Table S1:** Oxygen isotopic composition of chondrules.

| Name | n [a] | $\delta^{18}O$ | ±2σ | $\delta^{17}O$ | ±2σ | $\Delta^{17}O$ | ±2σ | $\Delta^{17}O$ (corr.)[b] | ±2σ | % relict olivines [c] |
|---|---|---|---|---|---|---|---|---|---|---|
| **OC (Ragland)** | | | | | | | | | | |
| MSc42 | 5 | 4.7 | 1.6 | 3.7 | 1.5 | 1.3 | 1.2 | | | |
| MSc43 | 4 | 5.3 | 1.4 | 3.6 | 1.2 | 0.8 | 0.6 | | | |
| MSc44 | 5 | 5.0 | 3.5 | 4.2 | 1.9 | 1.6 | 2.3 | | | |
| MSc45 | 4 | 4.9 | 1.3 | 4.4 | 0.8 | 1.8 | 1.0 | | | |
| MSc49 | 3 | 4.4 | 1.9 | 3.7 | 1.1 | 1.4 | 0.6 | | | |
| MSc51 | 5 | 4.0 | 0.8 | 3.9 | 0.7 | 1.8 | 0.8 | | | |
| MSc50 | 4 | 5.6 | 0.9 | 4.1 | 0.6 | 1.2 | 0.4 | | | |
| **Average** | | **4.8** | **1.1** | **3.9** | **0.6** | **1.4** | **0.7** | | | |
| **EC (MAC02837)** | | | | | | | | | | |
| MSc61 | 2 | 0.6 | 1.8 | 0.7 | 1.9 | 0.4 | 0.9 | | | |
| MSc55 | 6 | 2.6 | 3.0 | 1.2 | 1.8 | -0.2 | 0.5 | | | |
| MSc57 | 4 | 0.1 | 2.6 | 0.0 | 0.9 | 0.0 | 0.4 | | | |
| MSc58 | 6 | 1.8 | 1.6 | 0.9 | 1.4 | -0.1 | 0.8 | | | |
| **Average** | | **1.3** | **2.3** | **0.7** | **1.0** | **0.0** | **0.5** | | | |
| **CV (Allende)** | | | | | | | | | | |
| MSc2 | 13 | -8.7 | 1.5 | -9.7 | 1.2 | -5.2 | 1.3 | -5.2 | 1.3 | 0 |
| MSc3 | 10 | -5.8 | 1.8 | -9.0 | 0.8 | -5.9 | 0.8 | -5.9 | 0.8 | 0 |
| MSc5 | 5 | -4.9 | 1.5 | -7.4 | 1.5 | -4.8 | 1.2 | -4.8 | 1.2 | 0 |
| MSc8 | 6 | -2.6 | 2.1 | -4.3 | 2.0 | -3.0 | 1.2 | -3.0 | 1.2 | 0 |
| MSc10 | 20 | -8.2 | 11.6 | -10.6 | 11.4 | -6.4 | 5.4 | -5.0 | 1.9 | 30 |
| MSc12 | 18 | -6.7 | 6.7 | -9.3 | 5.9 | -5.9 | 2.8 | -5.4 | 1.4 | 17 |
| MSc14 | 10 | -10.9 | 9.6 | -12.9 | 8.5 | -7.2 | 3.7 | -5.7 | 1.1 | 50 |
| **Average** | | **-6.8** | **5.5** | **-9.0** | **5.4** | **-5.5** | **2.7** | **-5.0** | **2.0** | |
| **CR (GRA061000)** | | | | | | | | | | |
| MSc72 | 6 | -2.9 | 4.9 | -4.1 | 4.4 | -2.6 | 2.1 | -2.3 | 0.6 | 17 |
| MSc74 | 3 | -0.1 | 0.9 | -1.8 | 0.8 | -1.7 | 0.5 | -1.7 | 0.5 | 0 |
| MSc76 | 6 | -3.4 | 2.0 | -5.6 | 1.4 | -3.9 | 0.9 | -3.9 | 0.9 | 0 |
| MSc77 | 6 | 2.5 | 4.6 | -0.2 | 5.2 | -1.5 | 3.0 | -0.9 | 1.0 | 17 |
| MSc73 | 4 | -1.4 | 5.1 | -3.3 | 4.0 | -2.6 | 1.7 | -2.6 | 1.7 | 0 |
| MSc75 | 7 | -2.5 | 2.3 | -4.9 | 1.4 | -3.7 | 0.7 | -3.7 | 0.7 | 0 |
| **Average** | | **-1.3** | **4.3** | **-3.3** | **4.1** | **-2.7** | **1.0** | **-2.5** | **1.1** | |

[a] number of olivine grains measured
[b] relict grain corrected values for host chondrule olivines.
747 [c] relict olivines are by definition >3σ different from their host chodrule olivines (Ushikubo et al. 2012). The Percentage is given the simple abundances of relict

748

749

**Supplementary Materials**

**Fig. S1:** Images of the individual chondrules taken with a Keyence VHX-500F digital microscope (scale bar is 250 μm) and BSE images (scale bar as stated) of embedded chondrule fragments along with textural type (PP: porphyritic pyroxene; PO: porphyritic olivine; POP: porphyritic olivine-pyroxene; BO: barred olivine; RP: radial pyroxene) and FeO character (with type I having Fa< 10 and type II Fa >10).

Allende (CV3) chondrules:

MSc2

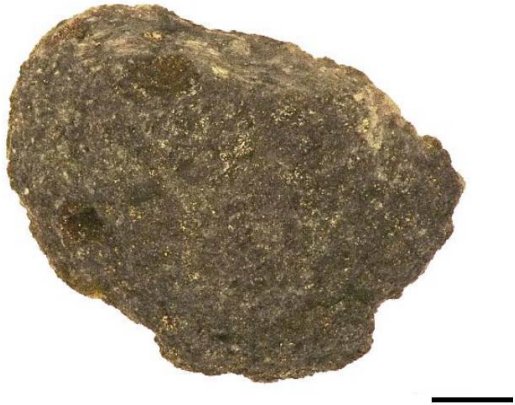

PO, Type I

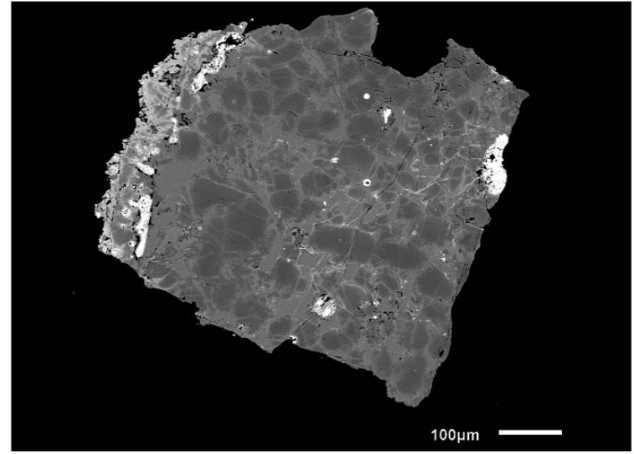

MSc3

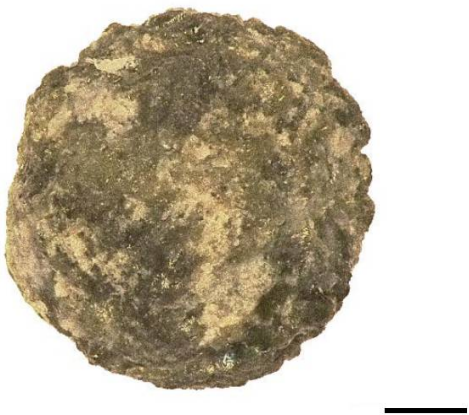

PO(P), Type I

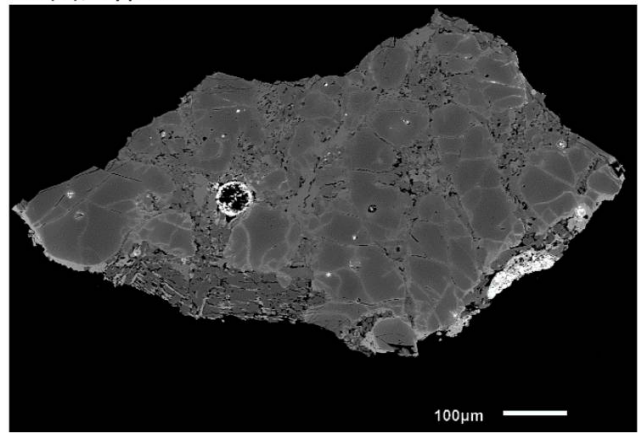

MSc5

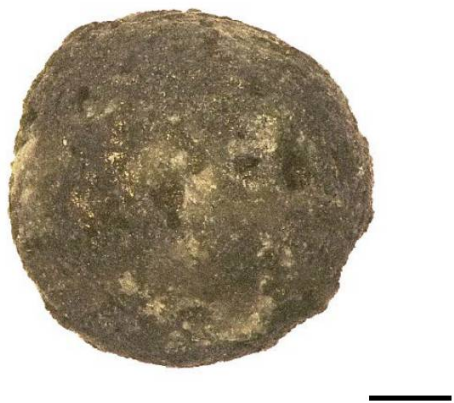

BO, Type I

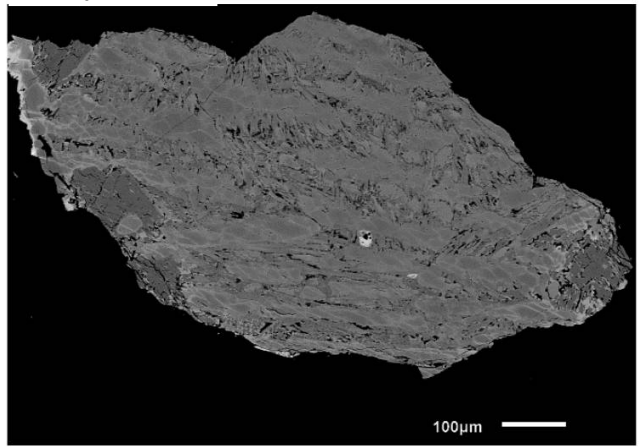

MSc8

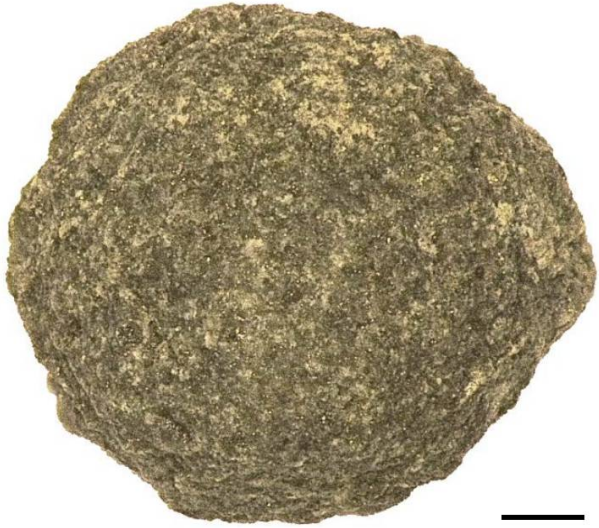

POP, Type I

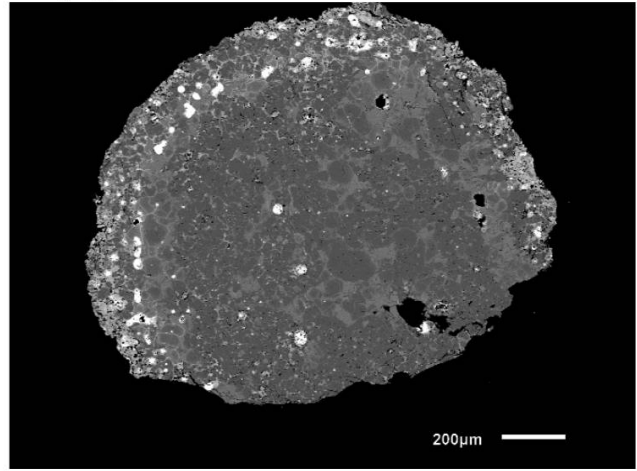

MSc10

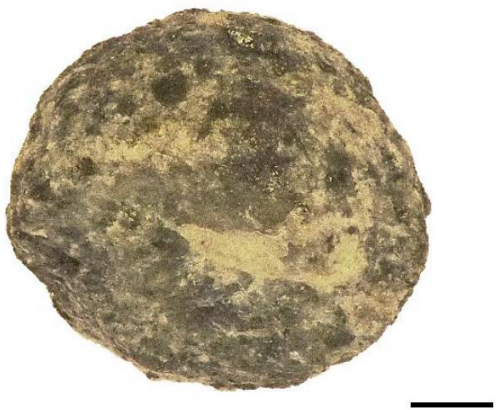

PO, Type I

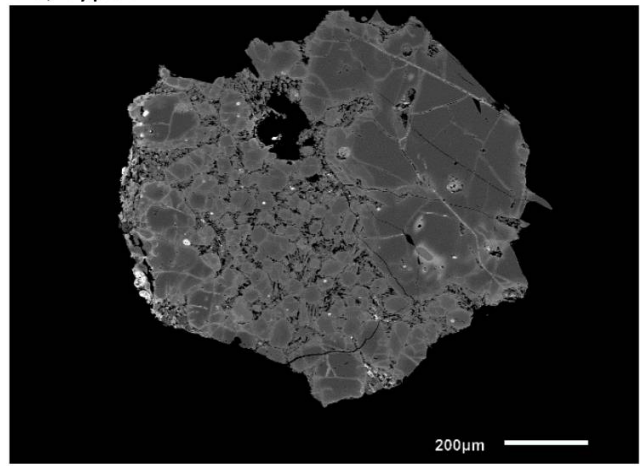

MSc12

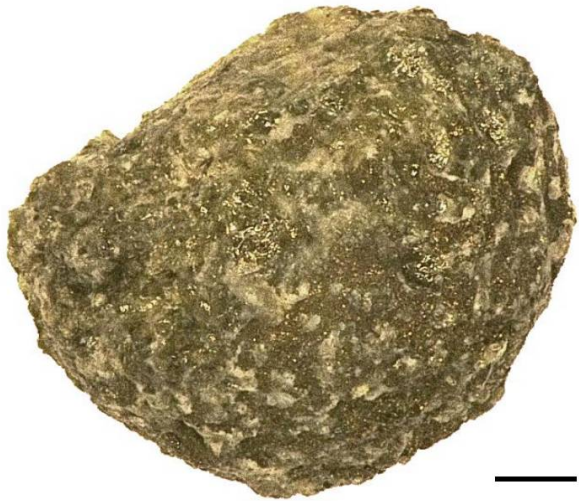

PO, Type I

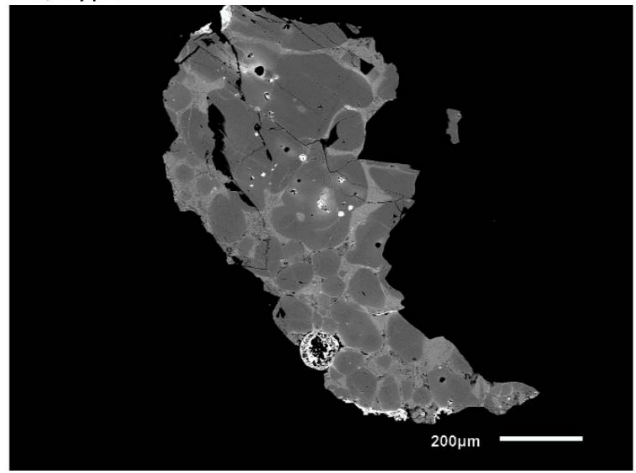

MSc14                      Al-rich, Type I

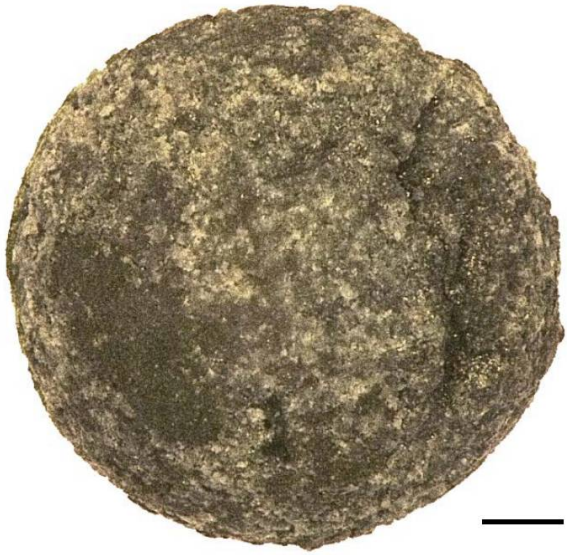
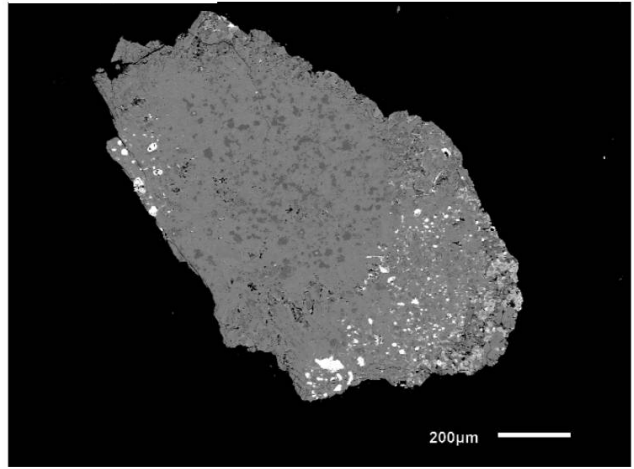

GRA06100 (CR2) chondrules:

MSc72

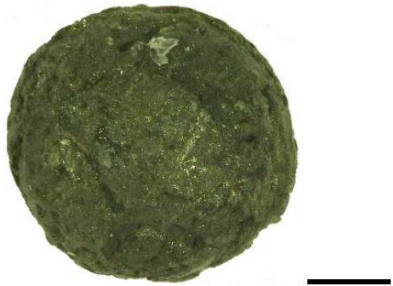

POP, Type I

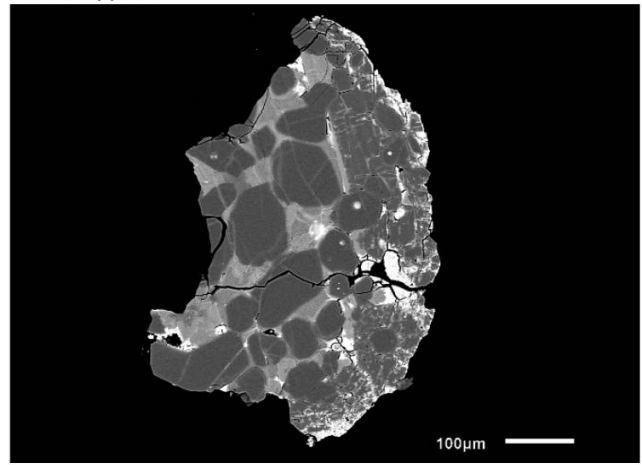

MSc73

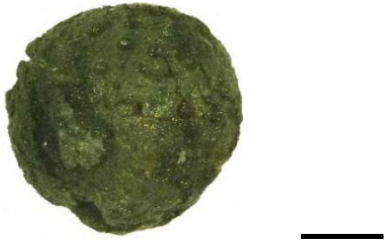

POP, Type I

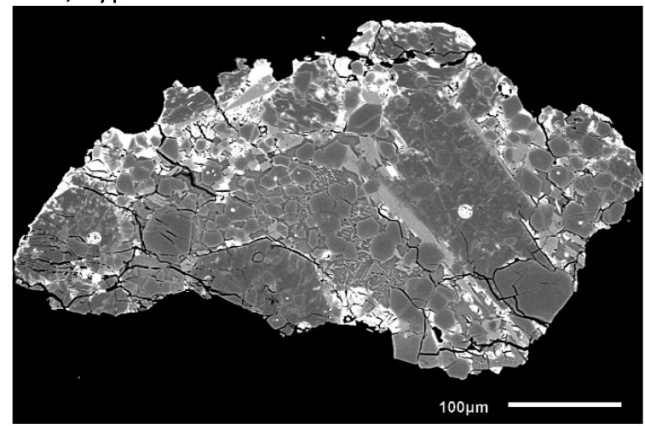

MSc74

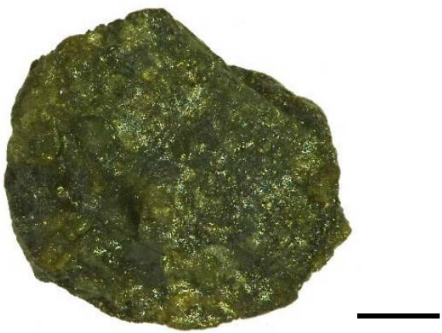

POP, Type I

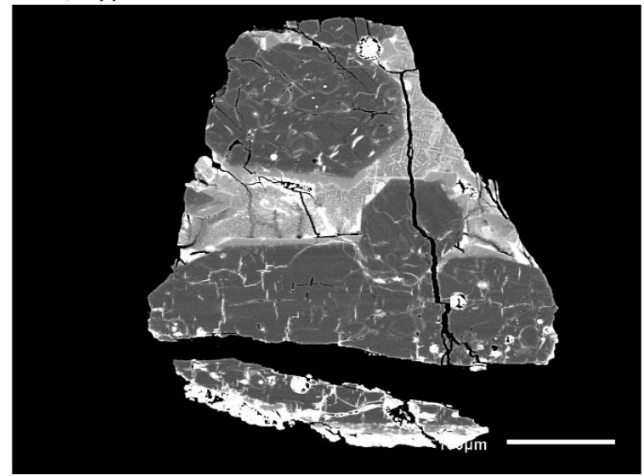

MSc75

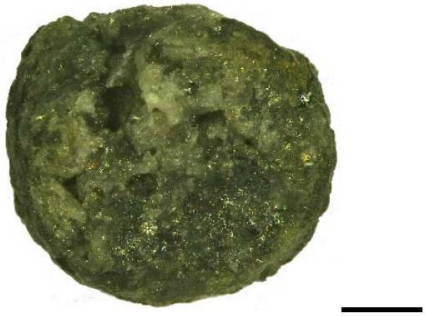

POP, Type I

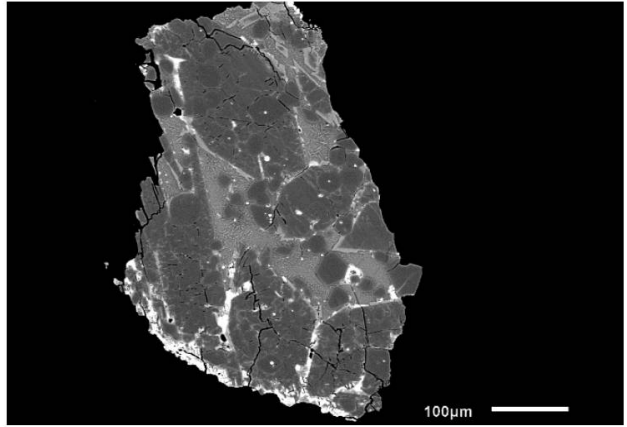

MSc76

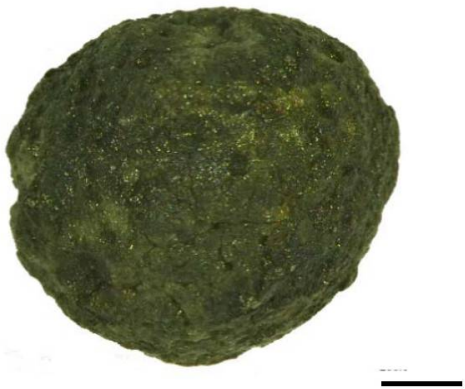

POP, Type I

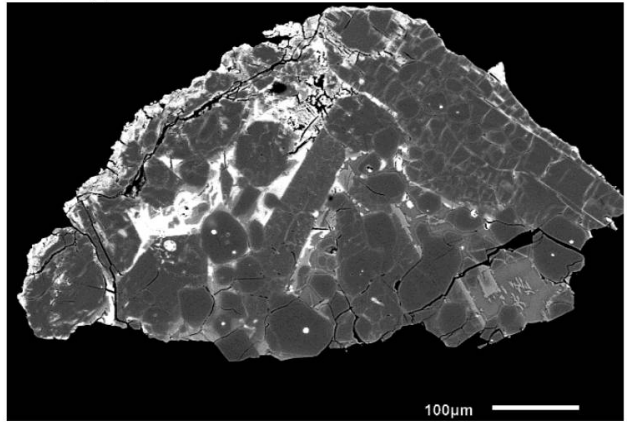

MSc77

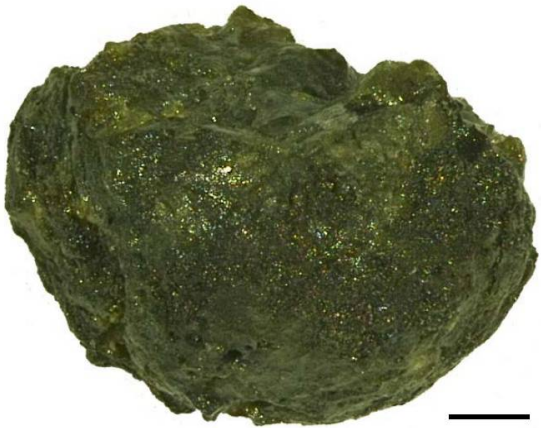

POP, Type I

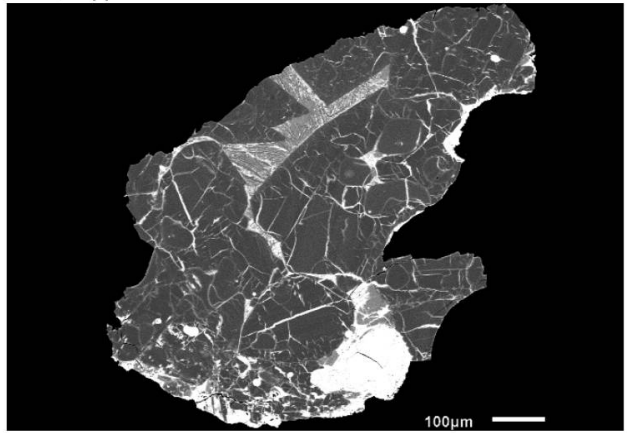

NWA081 (CR2) chondrule:

MSc78

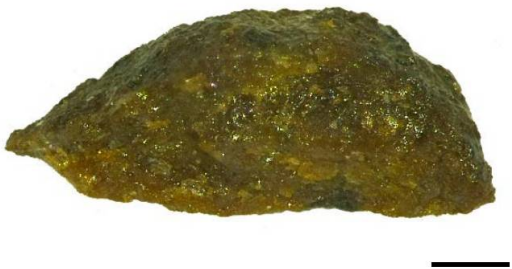

POP, Type I

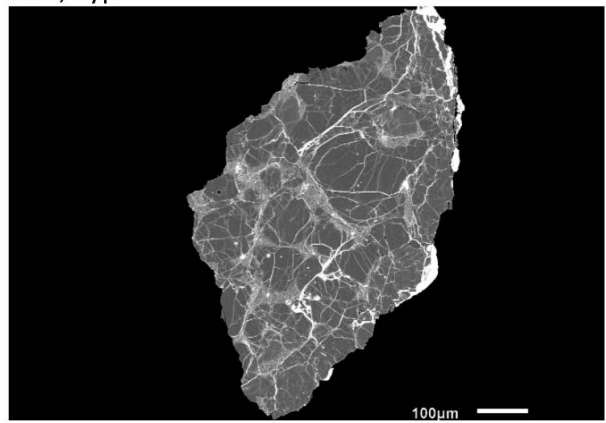

Ragland (LL3.4) chondrules:

MSc42

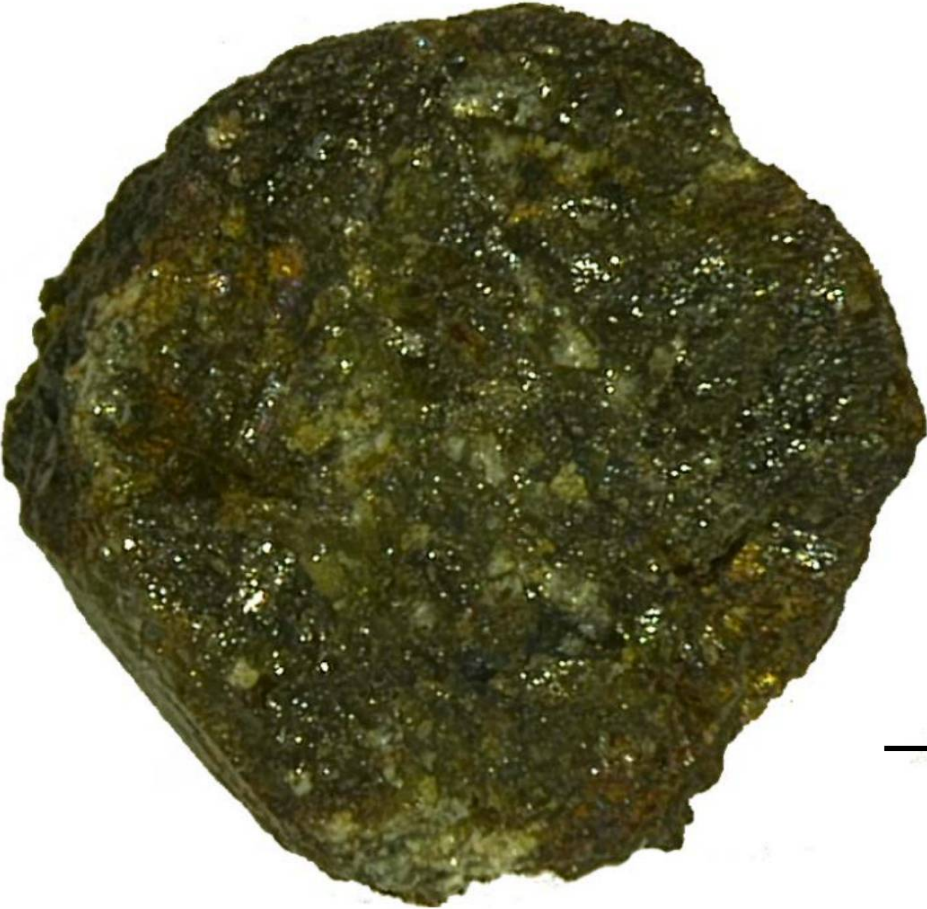

POP, Type II

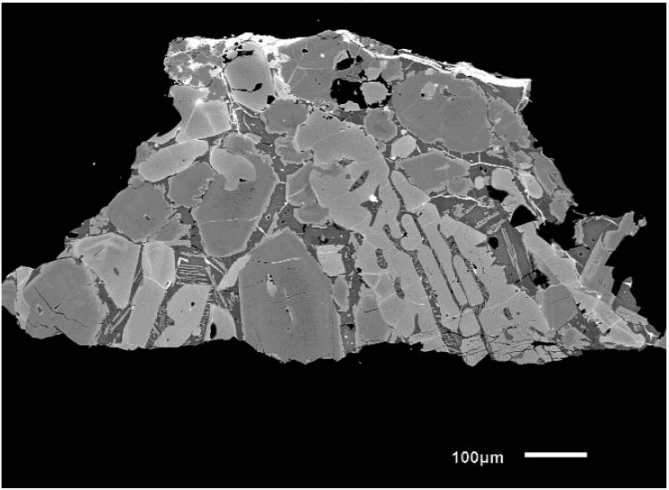

MSc43

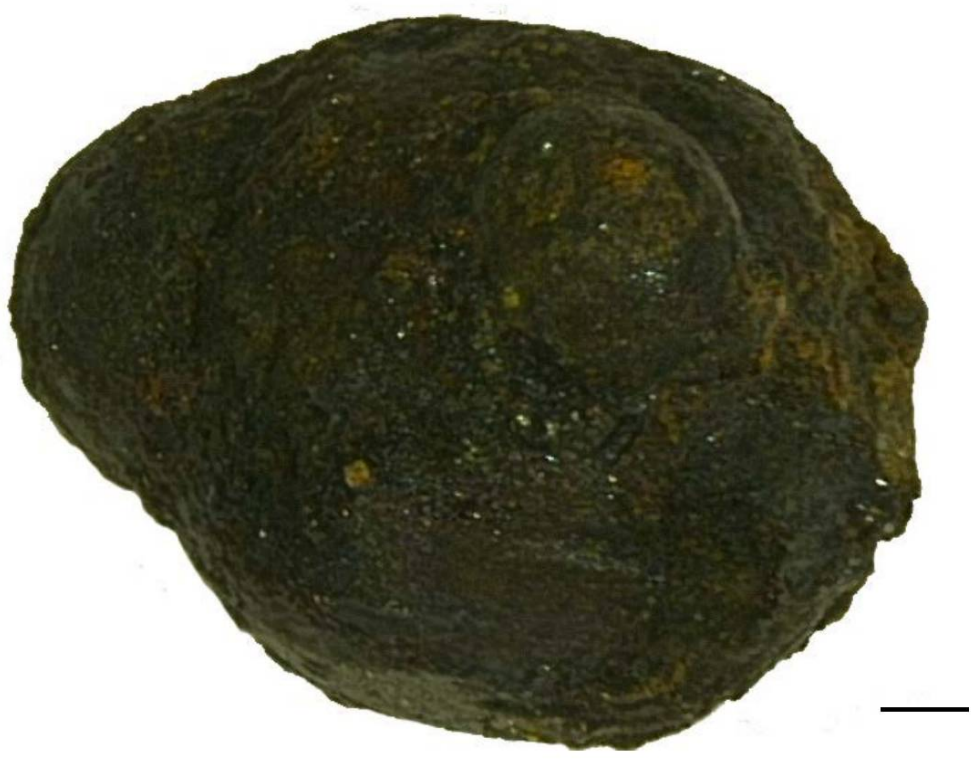

RP, Type II

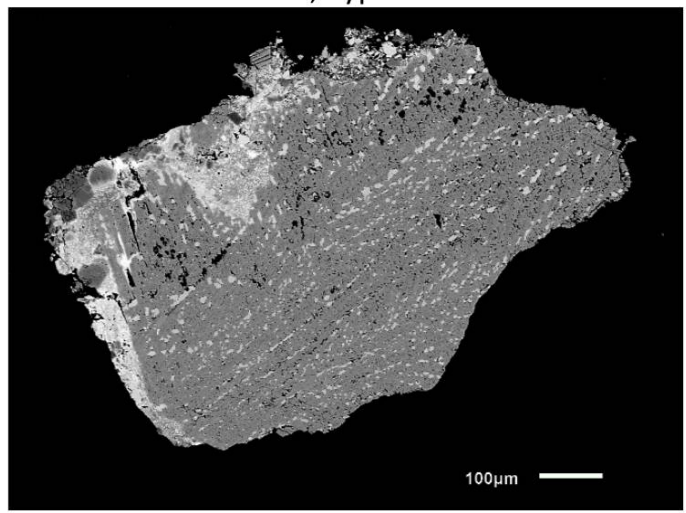

MSc44

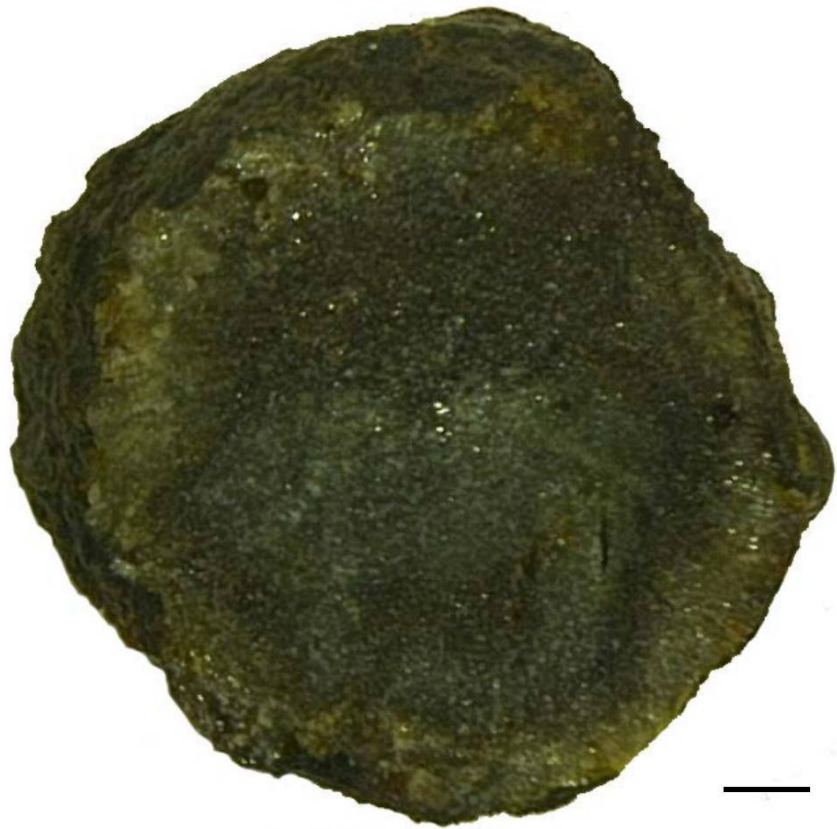

RP, Type II

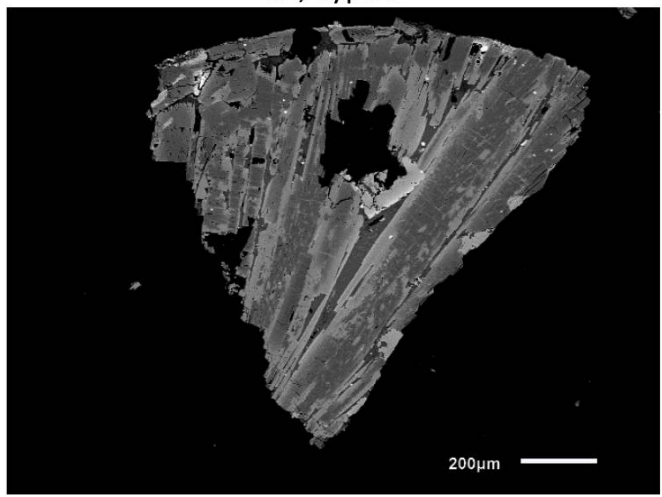

MSc45

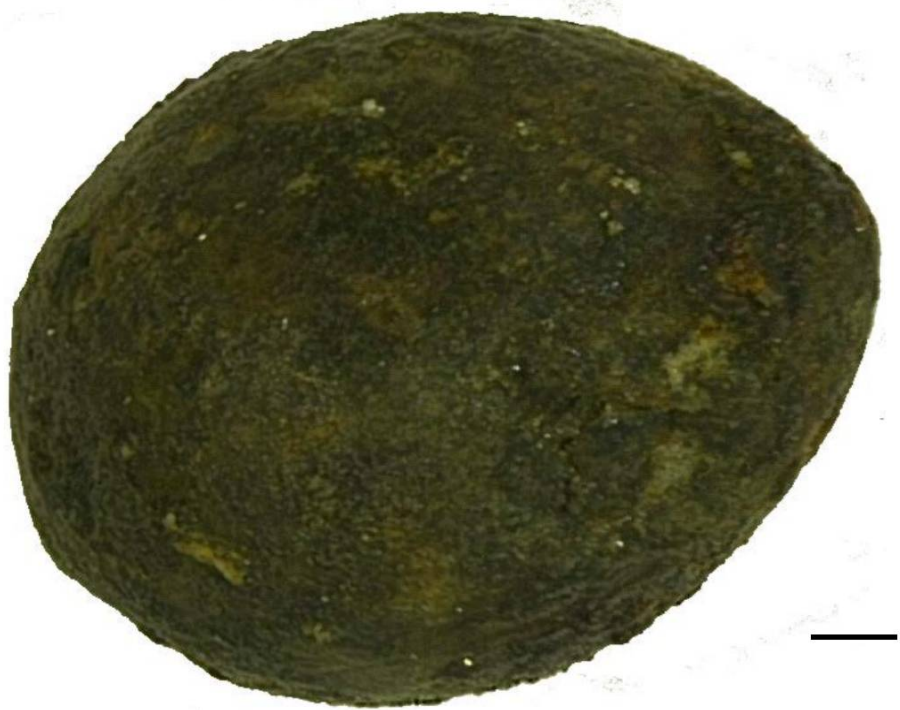

RP, Type II

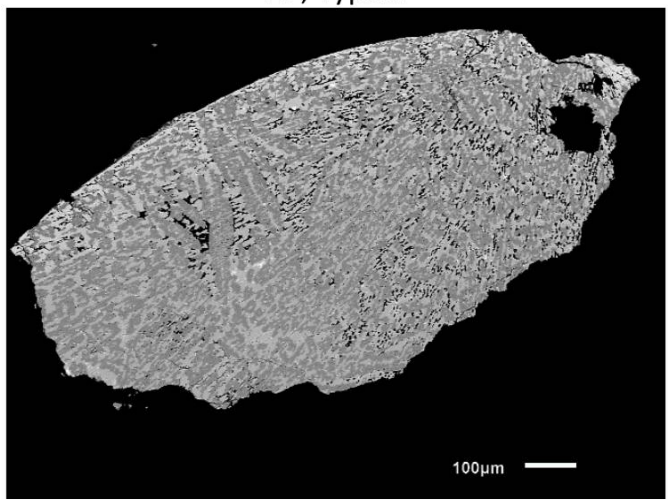

MSc49

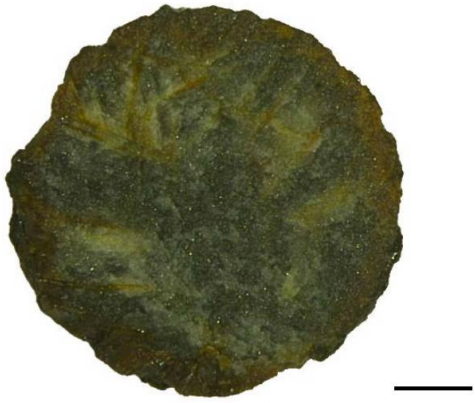

RP, Type II

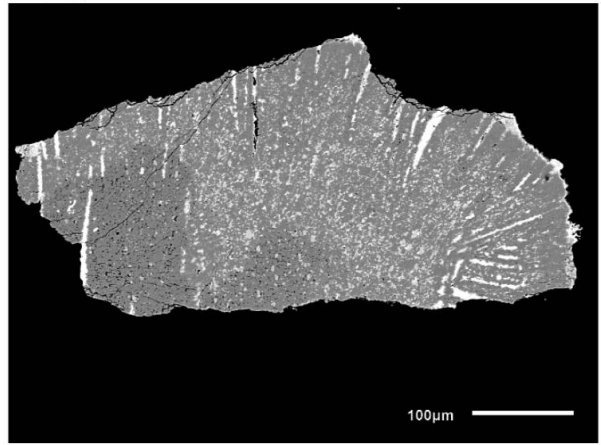

MSc50

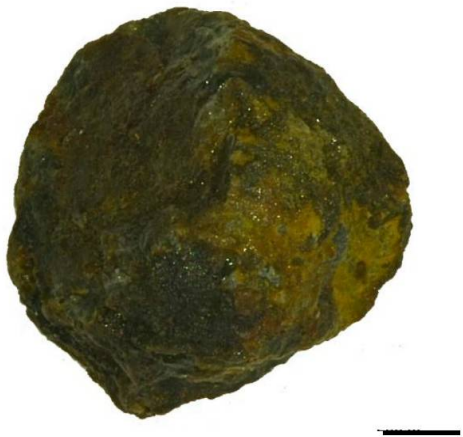

PO, Type II

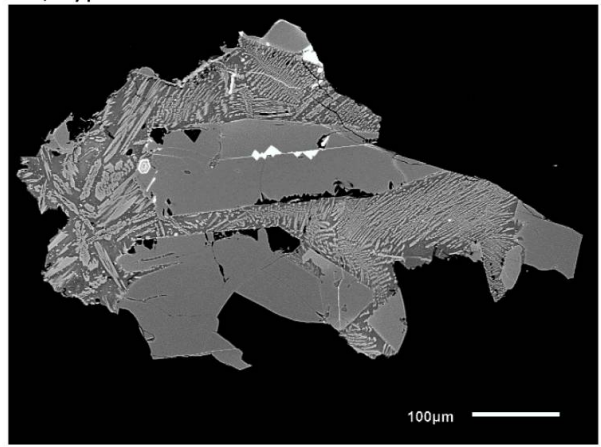

MSc51

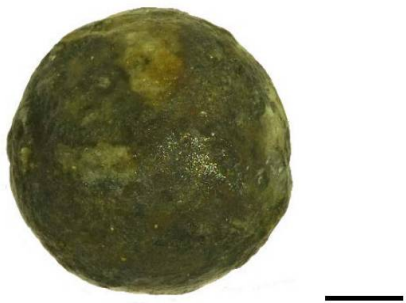

C, Type II

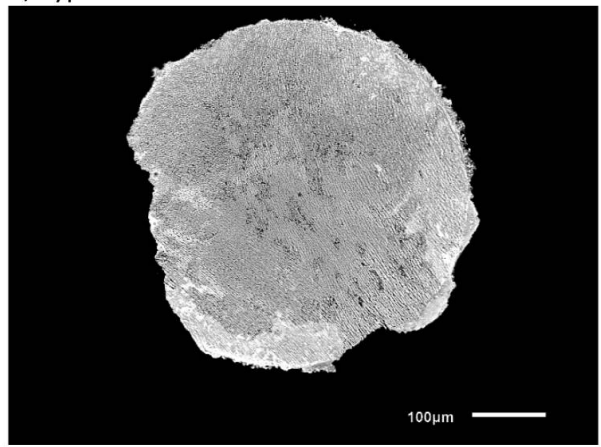

MSc52

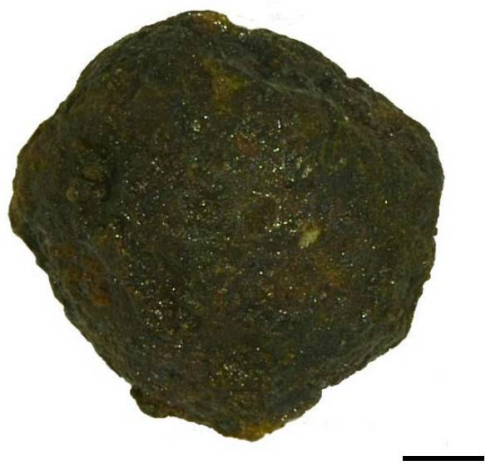

POP, Type I

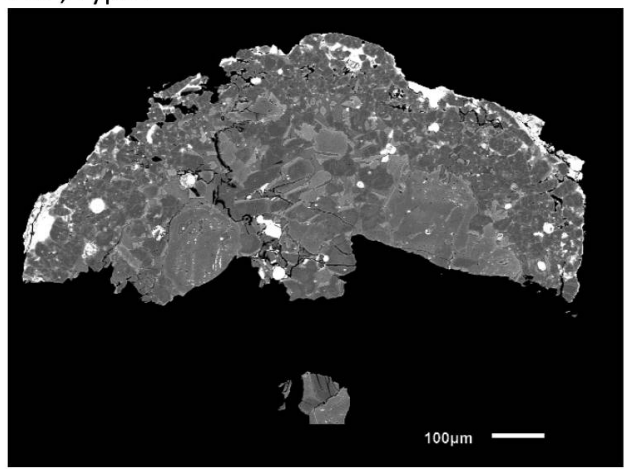

MSc53

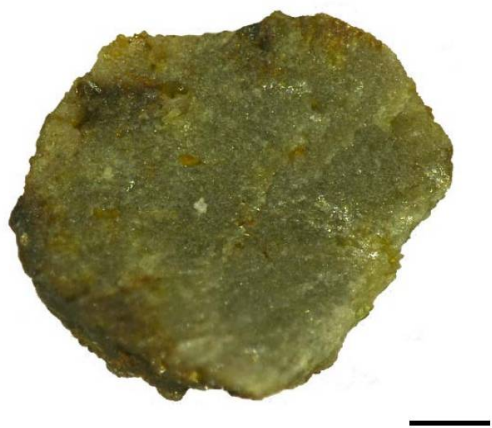

RP, Type II

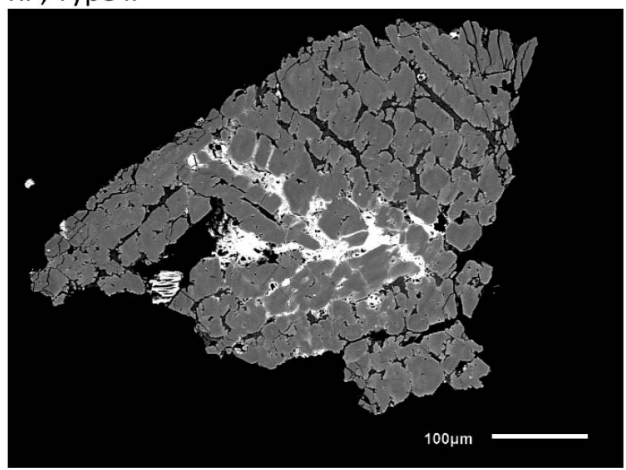

MAC02837 (EL3) chondrules:

MSc55 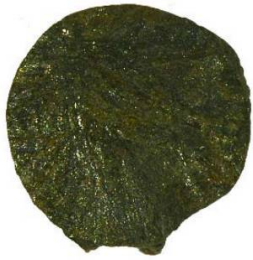

PP, Type I 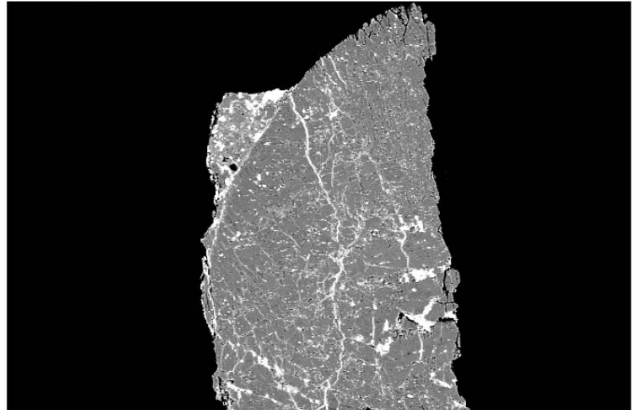

MSc57 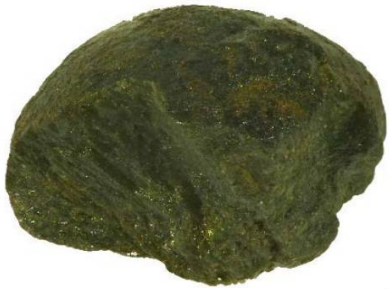

RP, Type I 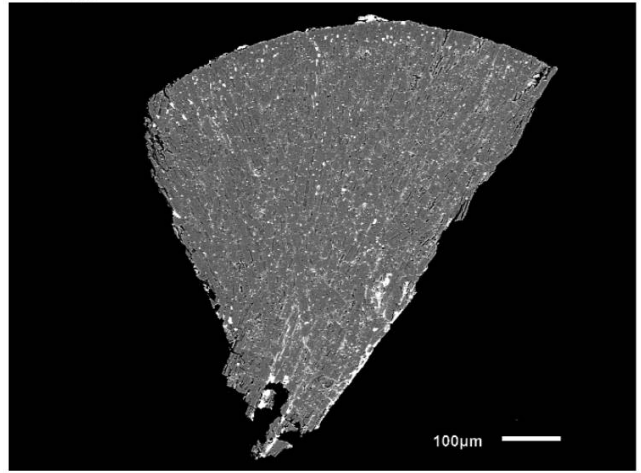

MSc58 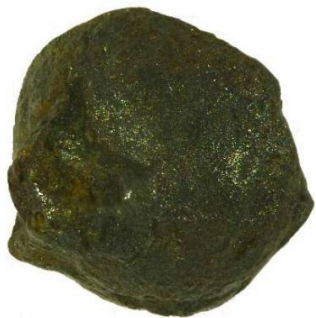

RP, Type I 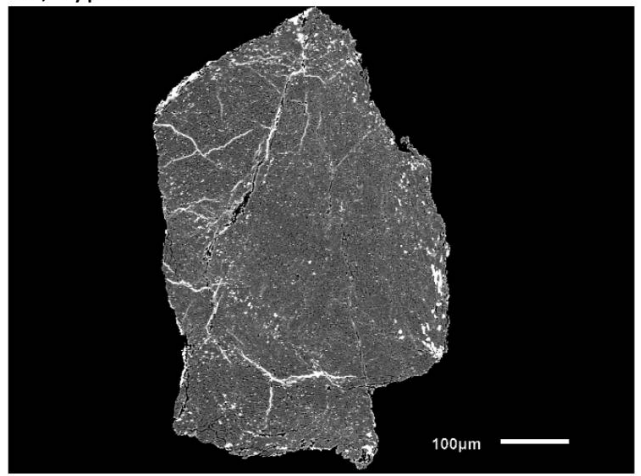

MSc61 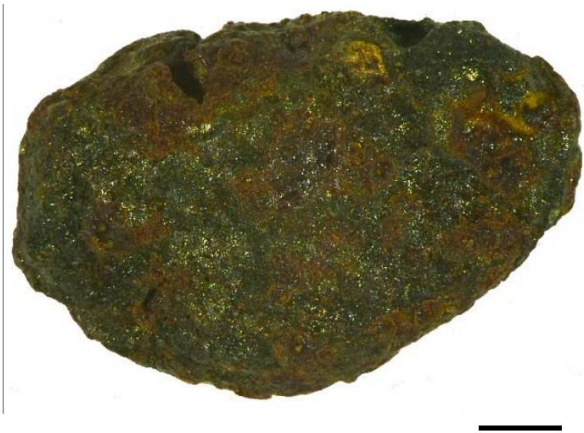

PP, Type I 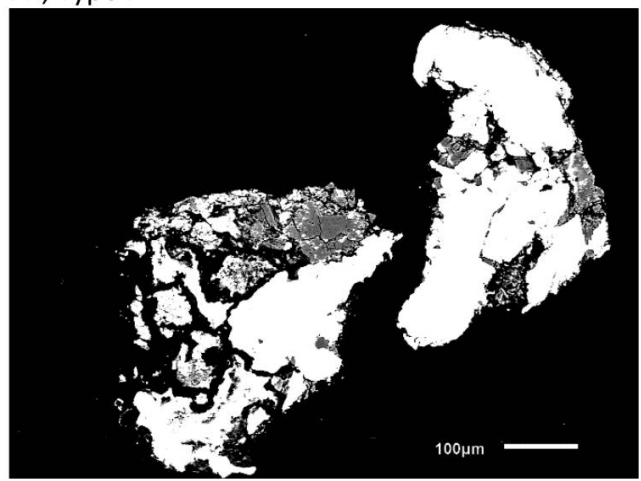

**Table S1:** Oxygen isotopic composition of chondrules.

| Name | n [a] | $\delta^{18}O$ | ±2σ | $\delta^{17}O$ | ±2σ | $\Delta^{17}O$ | ±2σ | $\Delta^{17}O$ (corr.)[b] | ±2σ | % relict olivines [c] |
|---|---|---|---|---|---|---|---|---|---|---|
| **OC (Ragland)** | | | | | | | | | | |
| MSc42 | 5 | 4.7 | 1.6 | 3.7 | 1.5 | 1.3 | 1.2 | | | |
| MSc43 | 4 | 5.3 | 1.4 | 3.6 | 1.2 | 0.8 | 0.6 | | | |
| MSc44 | 5 | 5.0 | 3.5 | 4.2 | 1.9 | 1.6 | 2.3 | | | |
| MSc45 | 4 | 4.9 | 1.3 | 4.4 | 0.8 | 1.8 | 1.0 | | | |
| MSc49 | 3 | 4.4 | 1.9 | 3.7 | 1.1 | 1.4 | 0.6 | | | |
| MSc51 | 5 | 4.0 | 0.8 | 3.9 | 0.7 | 1.8 | 0.8 | | | |
| MSc50 | 4 | 5.6 | 0.9 | 4.1 | 0.6 | 1.2 | 0.4 | | | |
| **Average** | | **4.8** | **1.1** | **3.9** | **0.6** | **1.4** | **0.7** | | | |
| **EC (MAC02837)** | | | | | | | | | | |
| MSc61 | 2 | 0.6 | 1.8 | 0.7 | 1.9 | 0.4 | 0.9 | | | |
| MSc55 | 6 | 2.6 | 3.0 | 1.2 | 1.8 | -0.2 | 0.5 | | | |
| MSc57 | 4 | 0.1 | 2.6 | 0.0 | 0.9 | 0.0 | 0.4 | | | |
| MSc58 | 6 | 1.8 | 1.6 | 0.9 | 1.4 | -0.1 | 0.8 | | | |
| **Average** | | **1.3** | **2.3** | **0.7** | **1.0** | **0.0** | **0.5** | | | |
| **CV (Allende)** | | | | | | | | | | |
| MSc2 | 13 | -8.7 | 1.5 | -9.7 | 1.2 | -5.2 | 1.3 | -5.2 | 1.3 | 0 |
| MSc3 | 10 | -5.8 | 1.8 | -9.0 | 0.8 | -5.9 | 0.8 | -5.9 | 0.8 | 0 |
| MSc5 | 5 | -4.9 | 1.5 | -7.4 | 1.5 | -4.8 | 1.2 | -4.8 | 1.2 | 0 |
| MSc8 | 6 | -2.6 | 2.1 | -4.3 | 2.0 | -3.0 | 1.2 | -3.0 | 1.2 | 0 |
| MSc10 | 20 | -8.2 | 11.6 | -10.6 | 11.4 | -6.4 | 5.4 | -5.0 | 1.9 | 30 |
| MSc12 | 18 | -6.7 | 6.7 | -9.3 | 5.9 | -5.9 | 2.8 | -5.4 | 1.4 | 17 |
| MSc14 | 10 | -10.9 | 9.6 | -12.9 | 8.5 | -7.2 | 3.7 | -5.7 | 1.1 | 50 |
| **Average** | | **-6.8** | **5.5** | **-9.0** | **5.4** | **-5.5** | **2.7** | **-5.0** | **2.0** | |
| **CR (GRA061000)** | | | | | | | | | | |
| MSc72 | 6 | -2.9 | 4.9 | -4.1 | 4.4 | -2.6 | 2.1 | -2.3 | 0.6 | 17 |
| MSc74 | 3 | -0.1 | 0.9 | -1.8 | 0.8 | -1.7 | 0.5 | -1.7 | 0.5 | 0 |
| MSc76 | 6 | -3.4 | 2.0 | -5.6 | 1.4 | -3.9 | 0.9 | -3.9 | 0.9 | 0 |
| MSc77 | 6 | 2.5 | 4.6 | -0.2 | 5.2 | -1.5 | 3.0 | -0.9 | 1.0 | 17 |
| MSc73 | 4 | -1.4 | 5.1 | -3.3 | 4.0 | -2.6 | 1.7 | -2.6 | 1.7 | 0 |
| MSc75 | 7 | -2.5 | 2.3 | -4.9 | 1.4 | -3.7 | 0.7 | -3.7 | 0.7 | 0 |
| **Average** | | **-1.3** | **4.3** | **-3.3** | **4.1** | **-2.7** | **1.0** | **-2.5** | **1.1** | |

[a] number of olivine grains measured
[b] relict grain corrected values for host chondrule olivines.
[c] relict olivines are by definition >3σ different from their host chodrule olivines (Ushikubo et al. 2012). The Percentage is given the simple abundances of relict

**Supplementary Materials**

**Fig. S1:** Images of the individual chondrules taken with a Keyence VHX-500F digital microscope (scale bar is 250 μm) and BSE images (scale bar as stated) of embedded chondrule fragments along with textural type (PP: porphyritic pyroxene; PO: porphyritic olivine; POP: porphyritic olivine-pyroxene; BO: barred olivine; RP: radial pyroxene) and FeO character (with type I having Fa< 10 and type II Fa >10).

Allende (CV3) chondrules:

MSc2 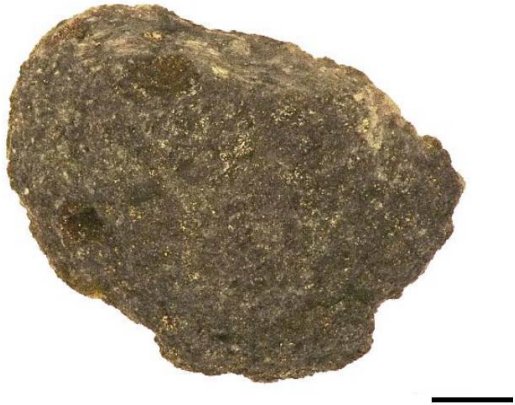 PO, Type I 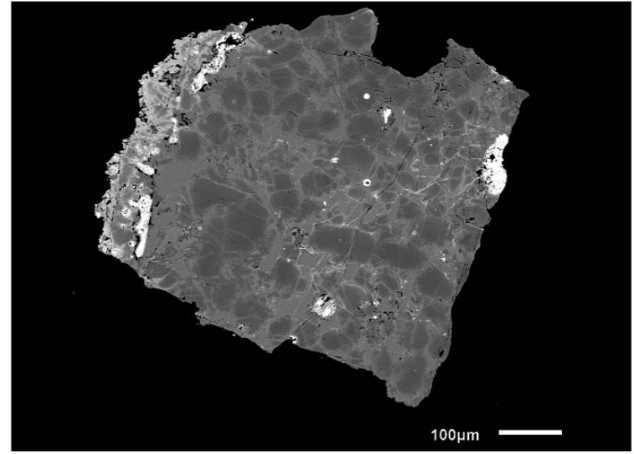

MSc3 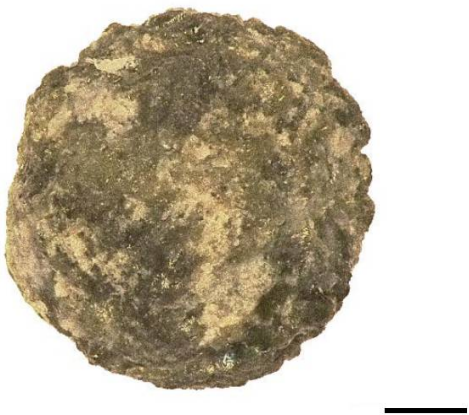 PO(P), Type I 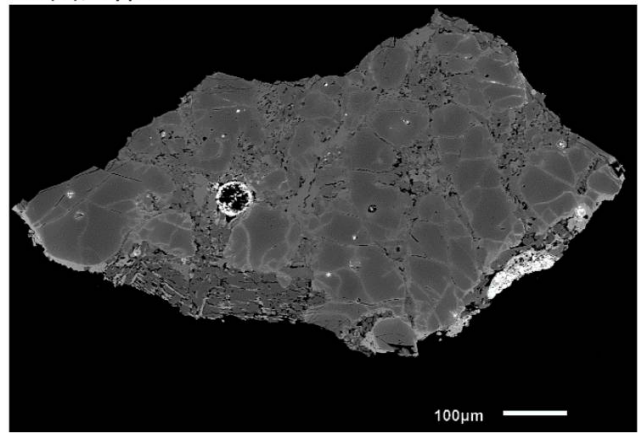

MSc5 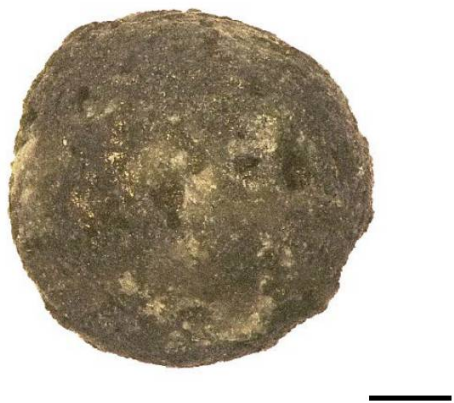 BO, Type I 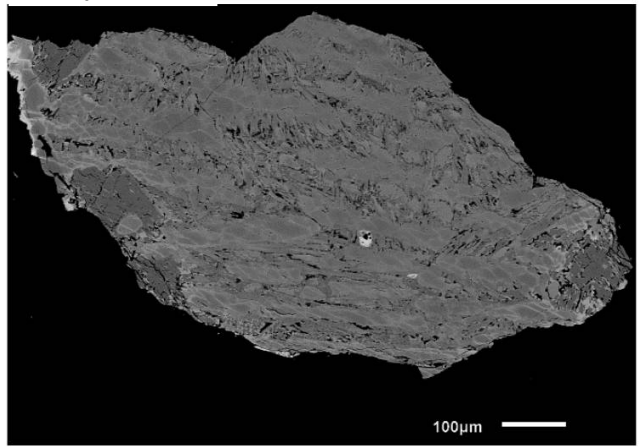

MSc8 POP, Type I

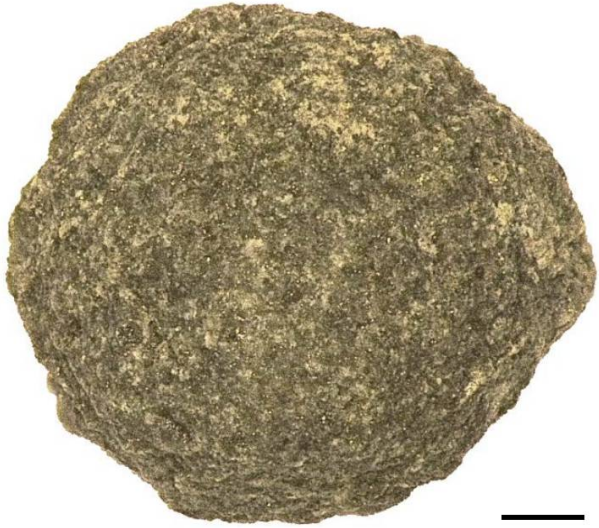
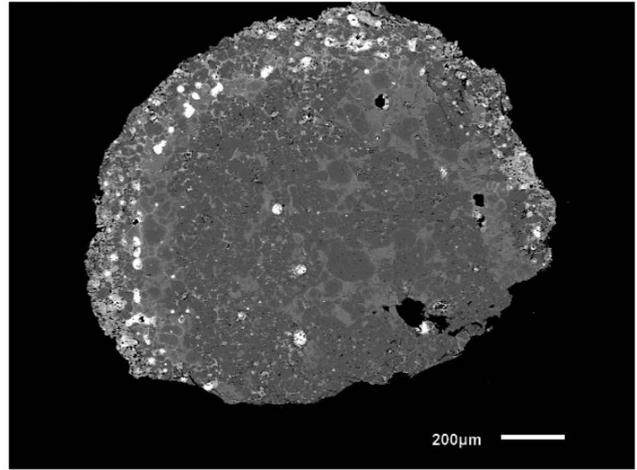

MSc10 PO, Type I

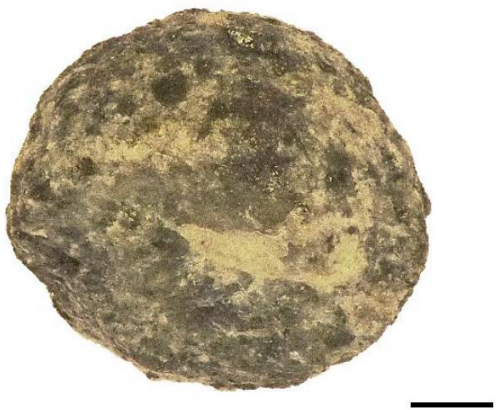
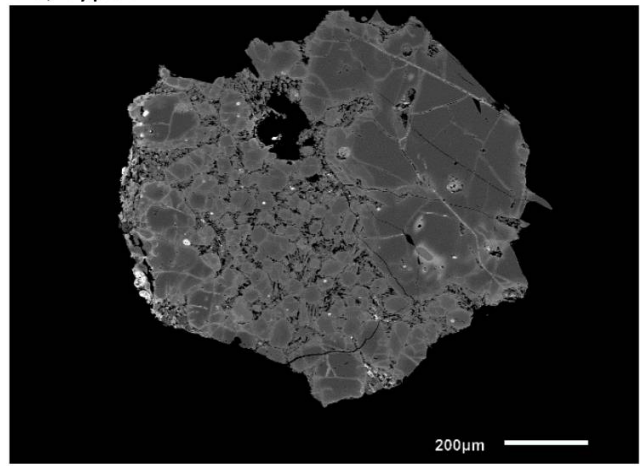

MSc12 PO, Type I

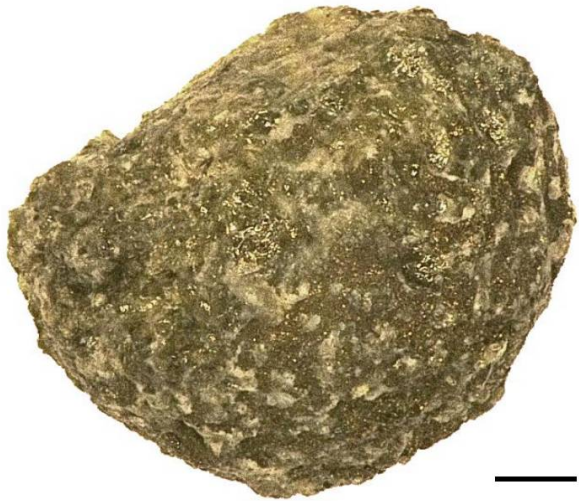
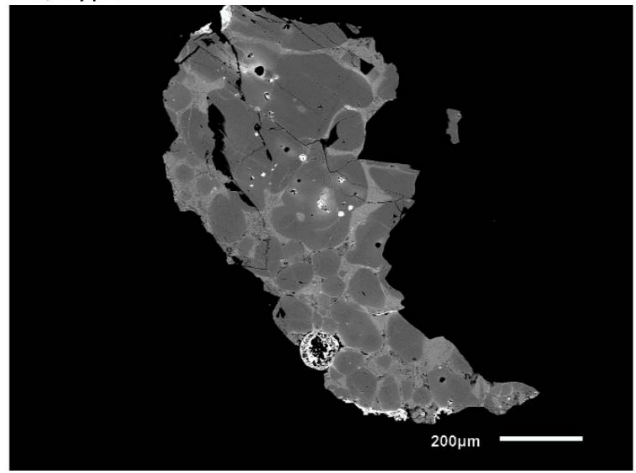

MSc14 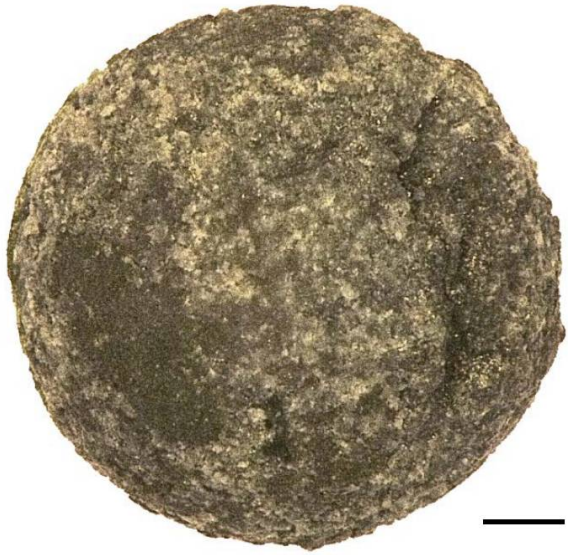

Al-rich, Type I 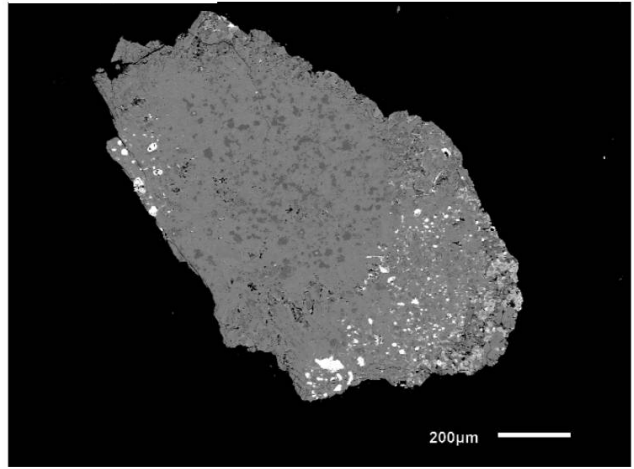

GRA06100 (CR2) chondrules:

MSc72            POP, Type I

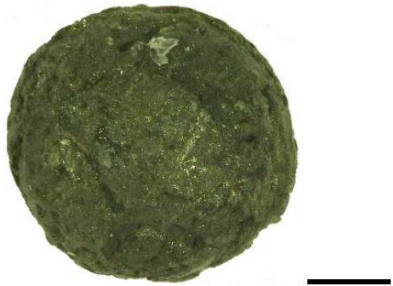
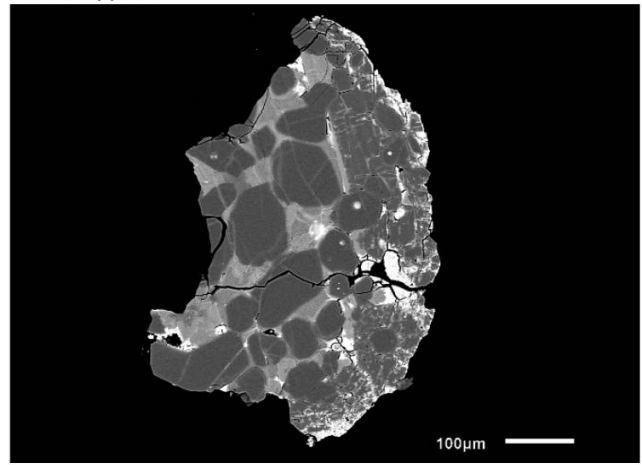

MSc73            POP, Type I

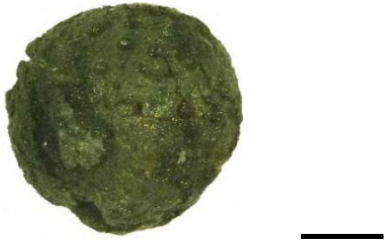
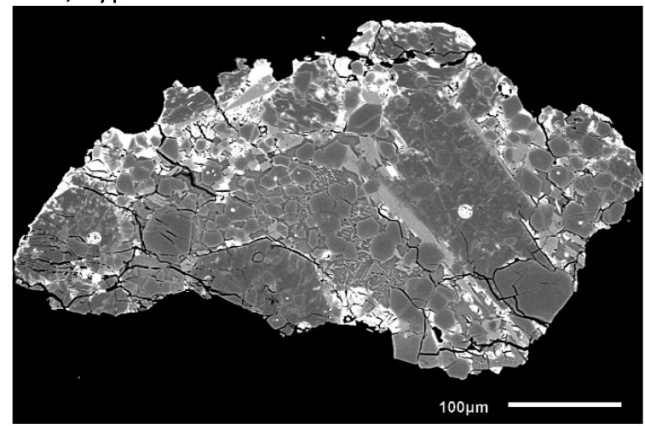

MSc74            POP, Type I

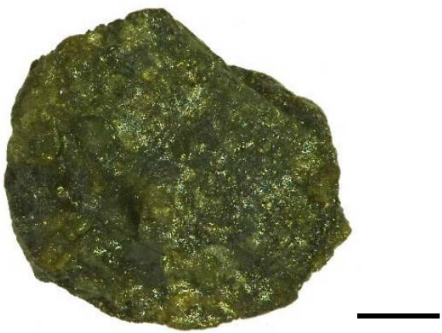
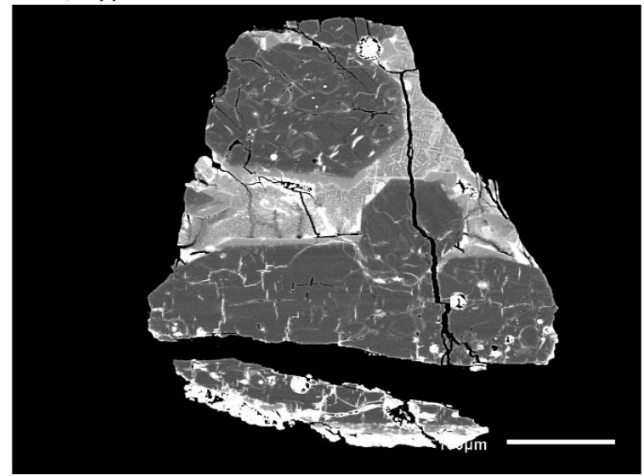

MSc75

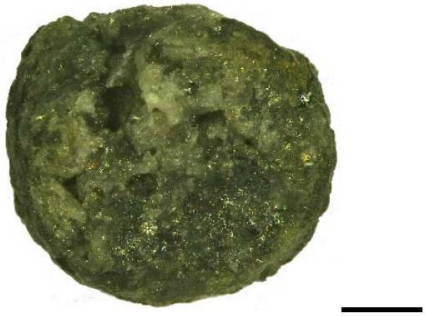

POP, Type I

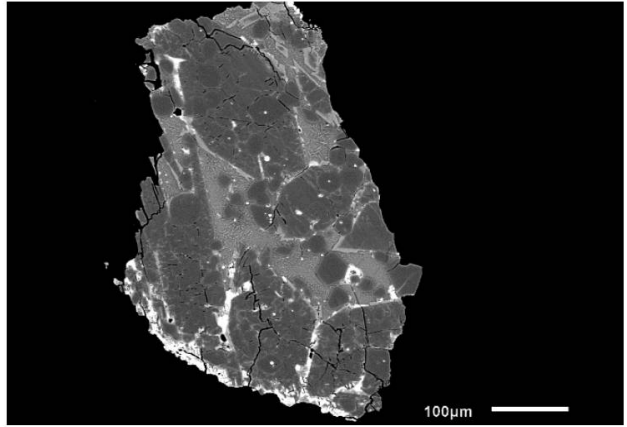

MSc76

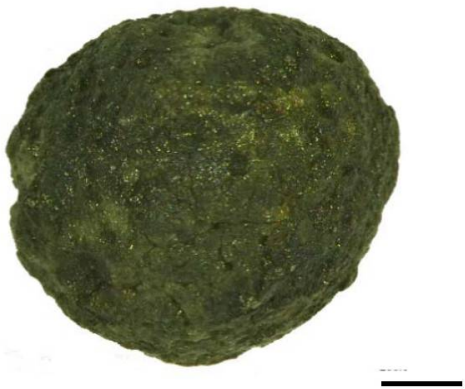

POP, Type I

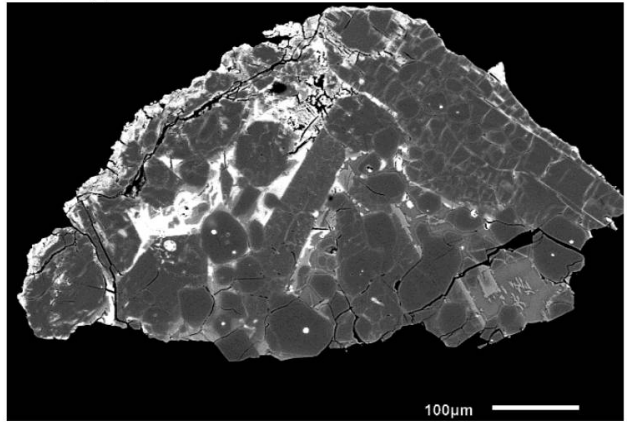

MSc77

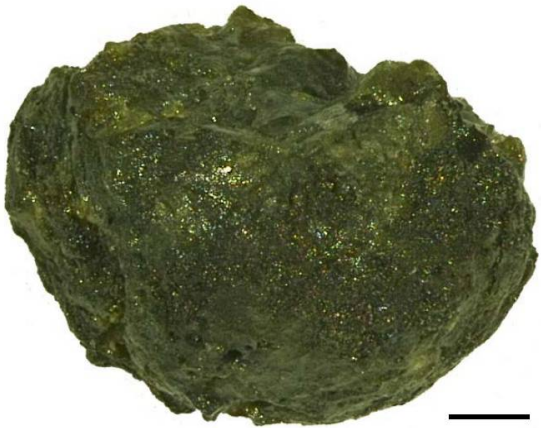

POP, Type I

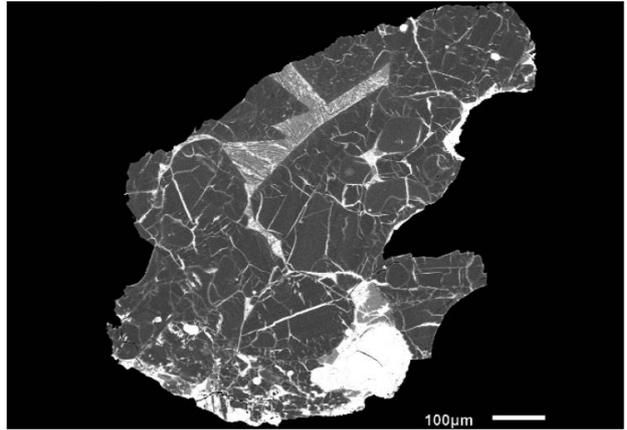

NWA081 (CR2) chondrule:

MSc78

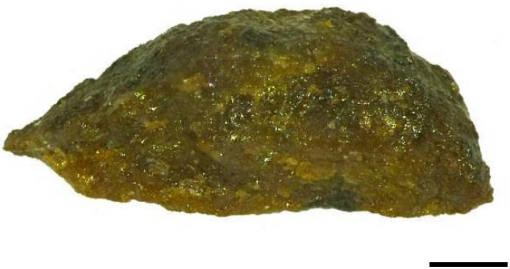

POP, Type I

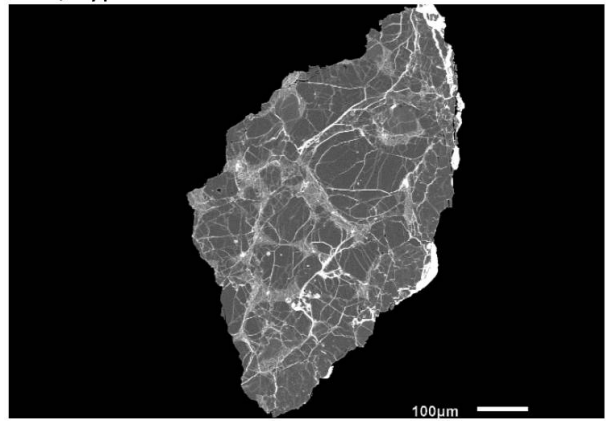

Ragland (LL3.4) chondrules:

MSc42

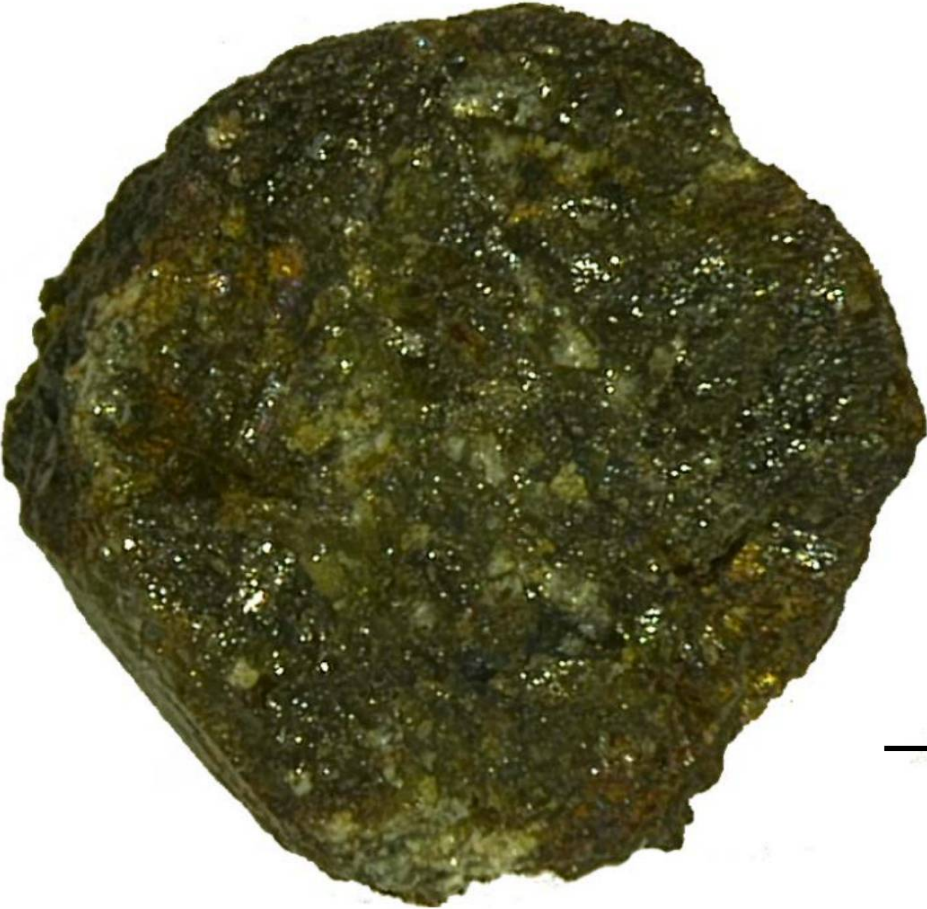

POP, Type II

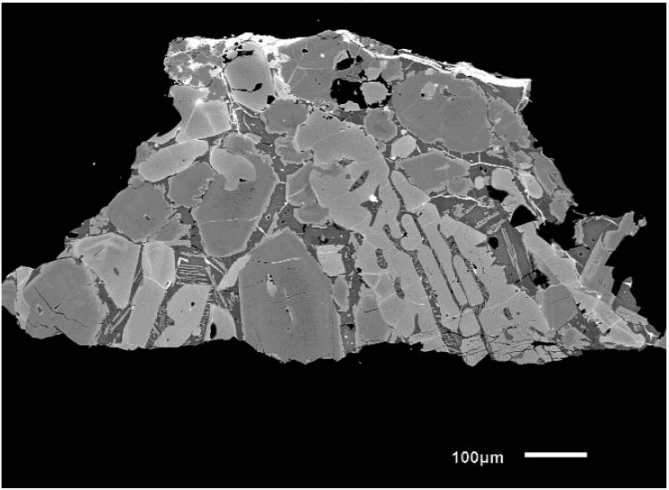

MSc43

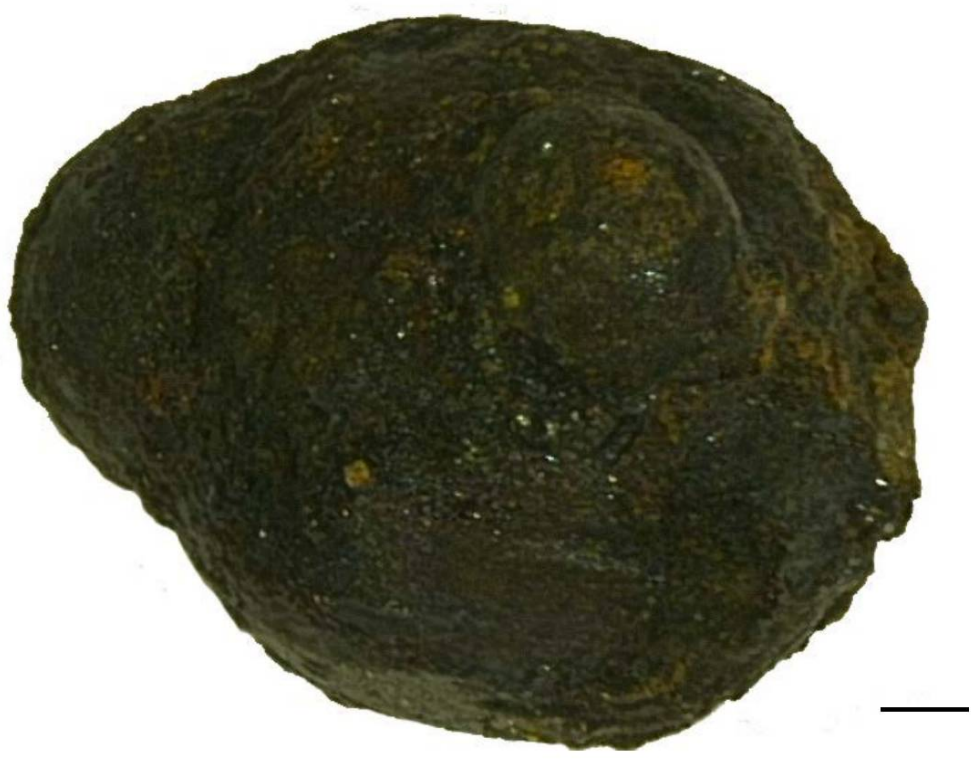

RP, Type II

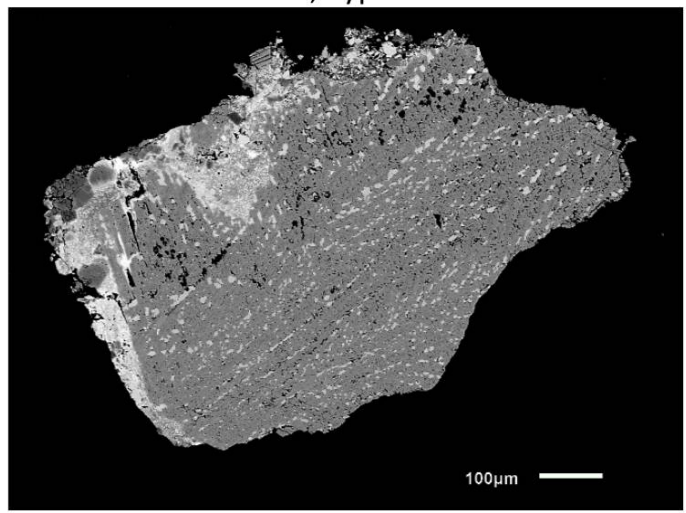

MSc44

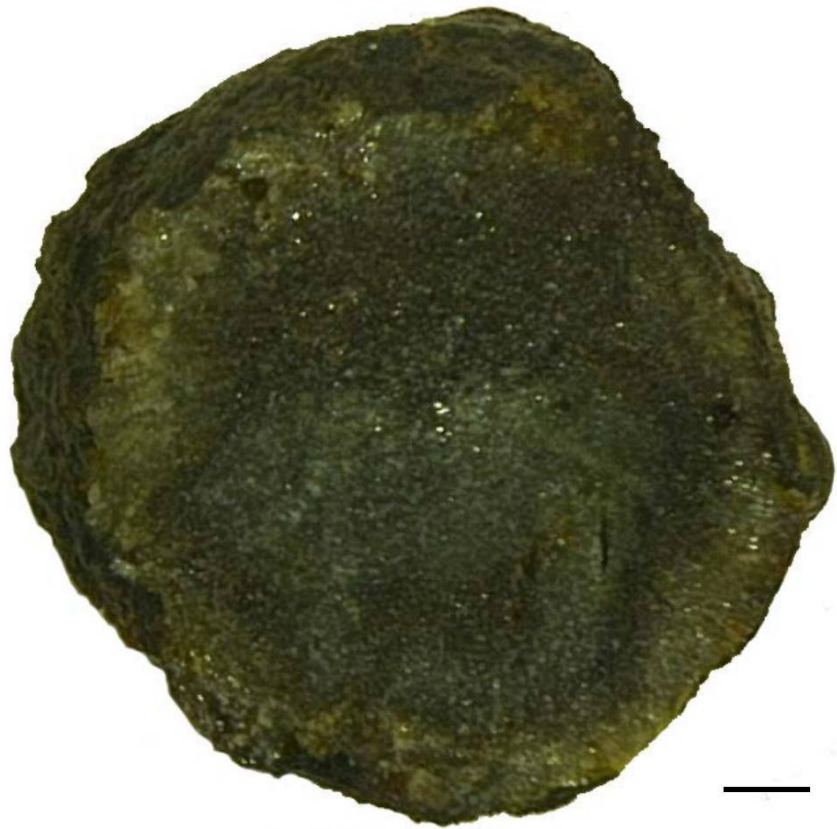

RP, Type II

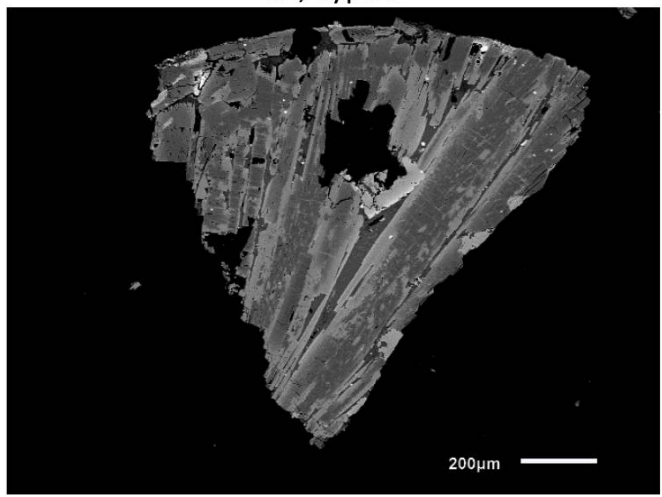

MSc45

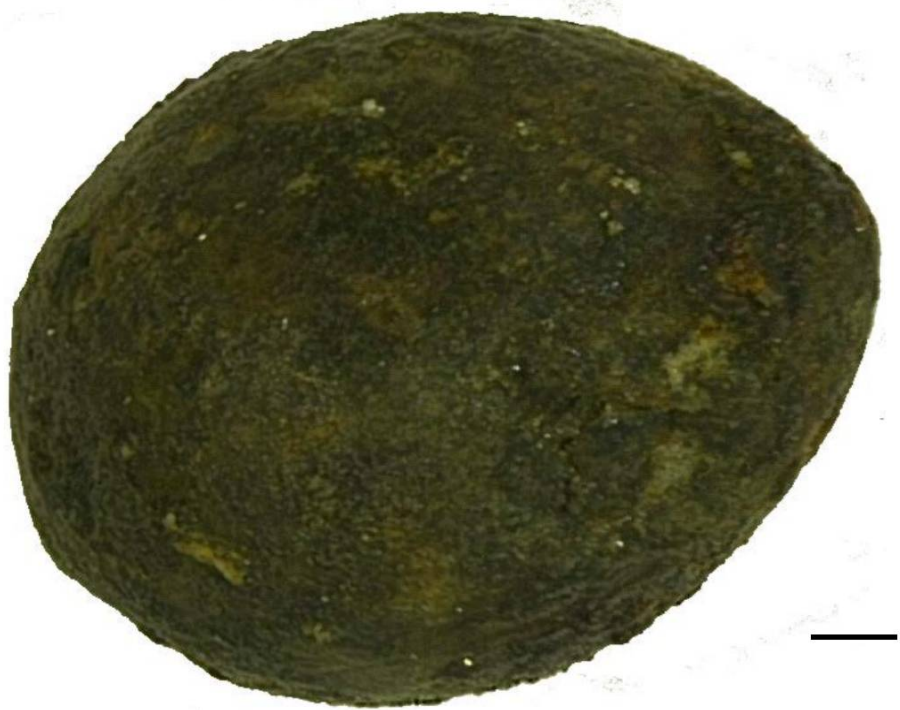

RP, Type II

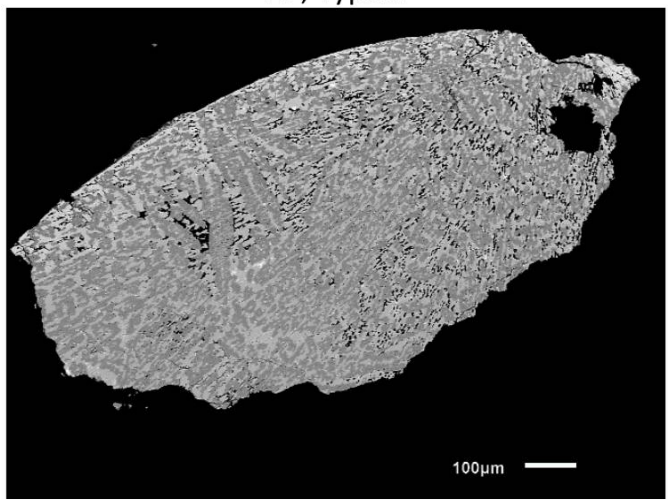

100μm

MSc49

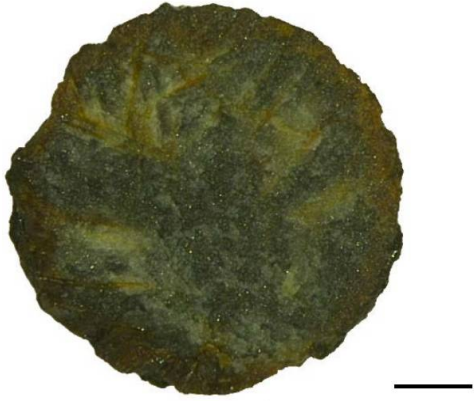

RP, Type II

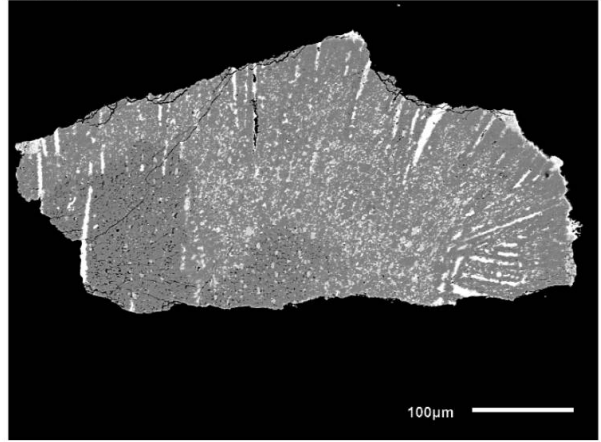

MSc50

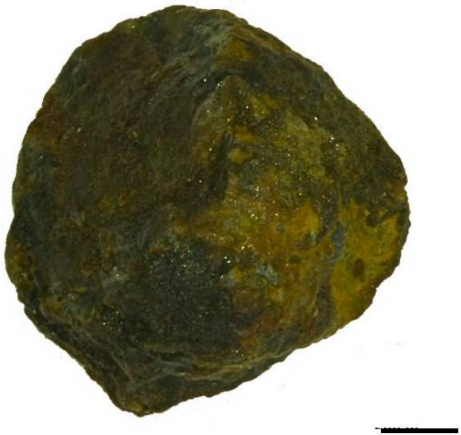

PO, Type II

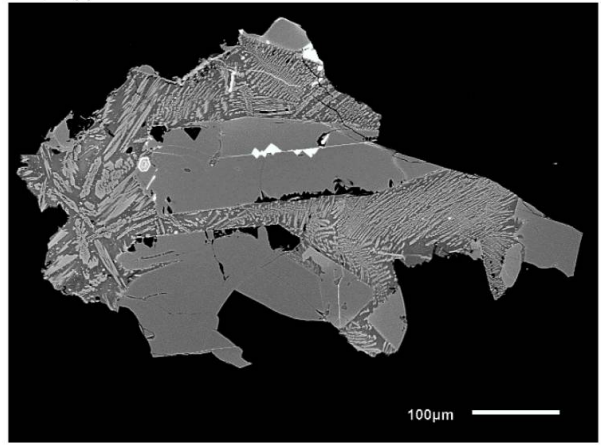

MSc51

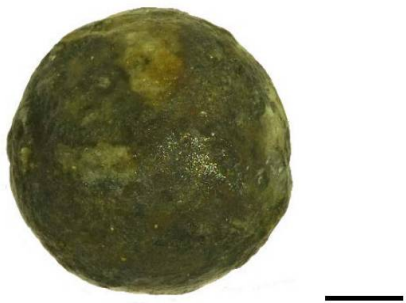

C, Type II

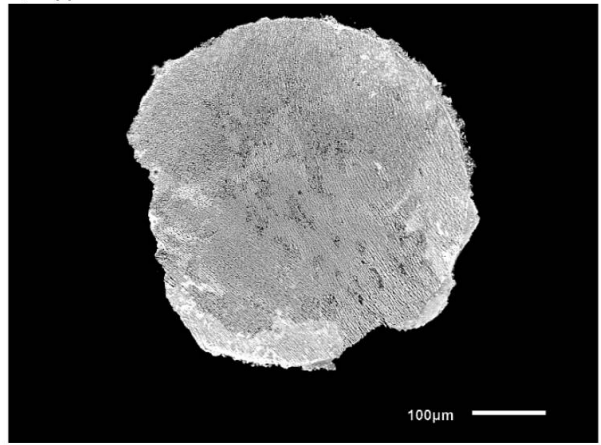

MSc52

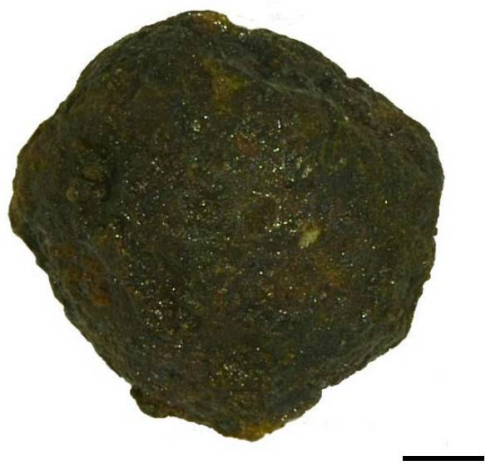

POP, Type I

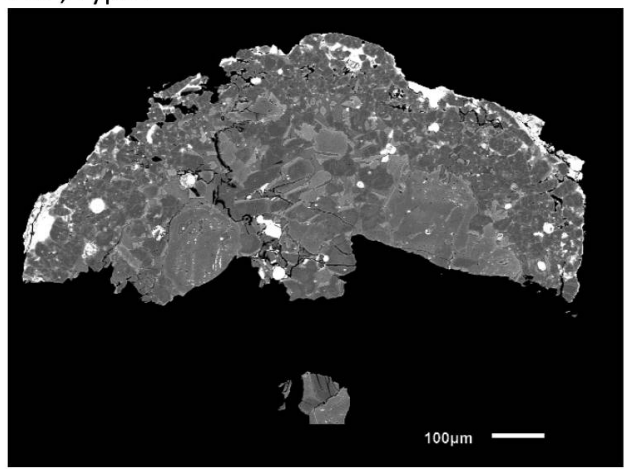

MSc53

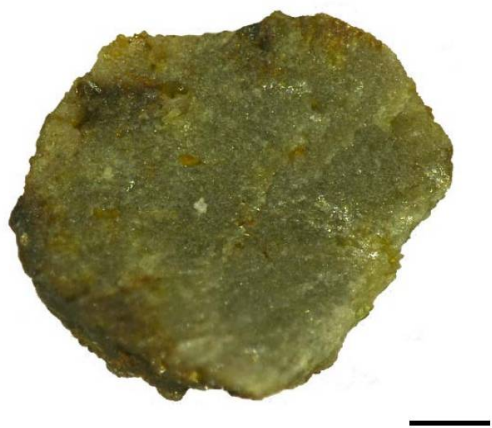

RP, Type II

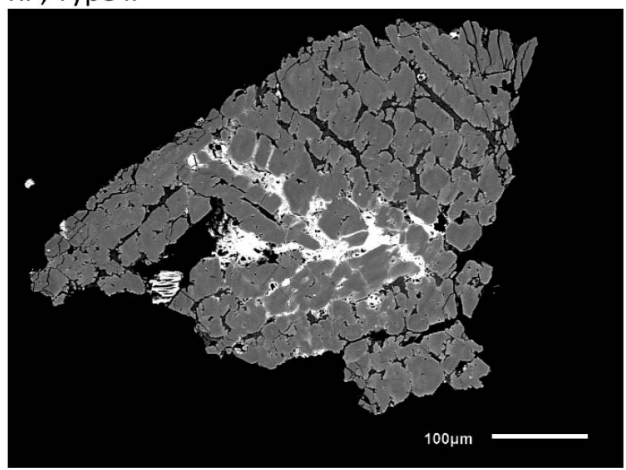

MAC02837 (EL3) chondrules:

MSc55  PP, Type I

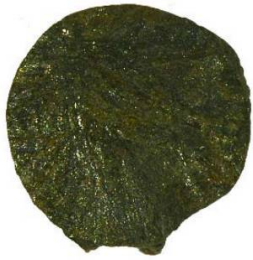
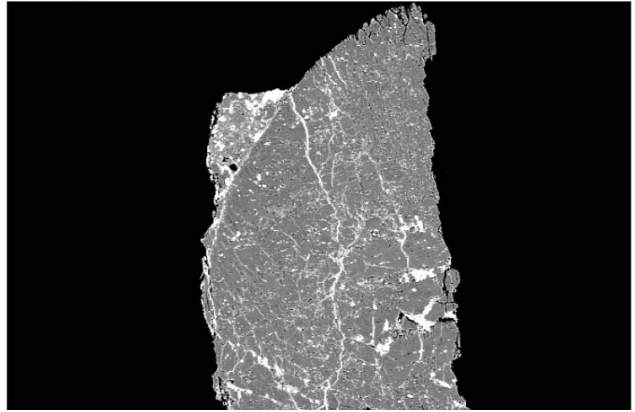

MSc57  RP, Type I

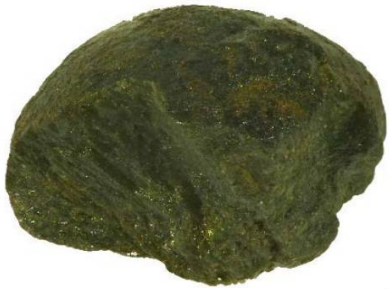
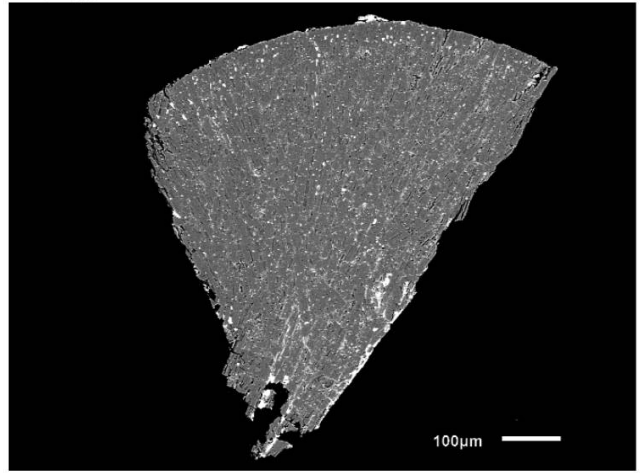

MSc58  RP, Type I

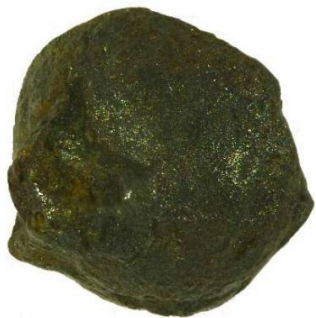
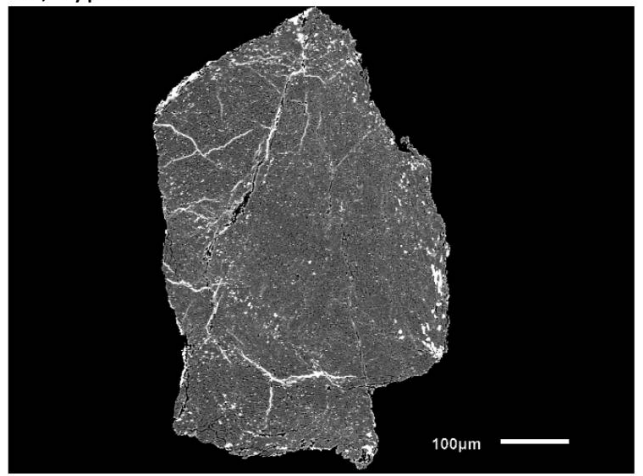

MSc61  PP, Type I

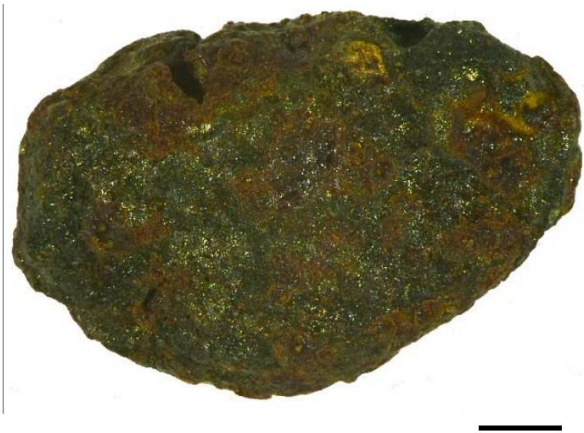
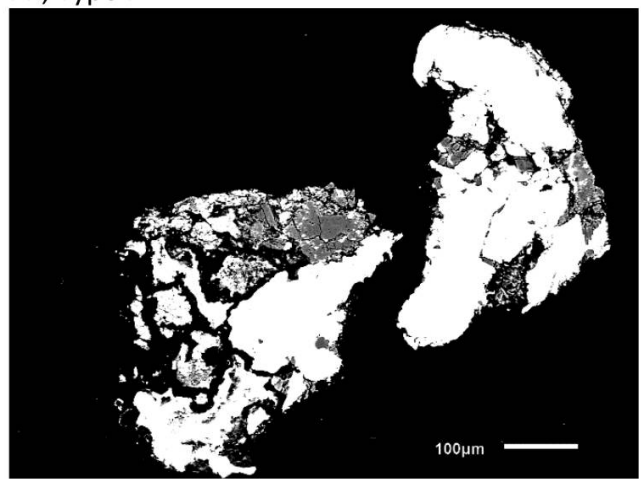